# Charge transfer on the metallic atom-pair bond, and the crystal structures adopted by intermetallic compounds

T. Rajasekharan [a)]
*Defence Metallurgical Research Laboratory, Kanchanbagh P.O. Hyderabad 500 058, India*
V. Seshubai
*School of Physics, University of Hyderabad, Hyderabad 500 046, India.*

**ABSTRACT**

Pauling had proposed that charge is transferred from the more electronegative atom to the less electronegative one during metallic alloying. An expression has recently been derived for the energy of the unlike atom-pair bond using that concept. In this paper, Pauling's view is further supported by showing that the electronegativity and valence differences between unlike atoms in binary metallic alloys are of the required magnitude and direction so as to bring the metallic radii ratio $R_A/R_B$ by charge transfer to a value $r_A/r_B$ for ideal packing in an intermetallic compound. It is also demonstrated that lattice parameters of intermetallic compounds belonging to several crystal structure types can be obtained with excellent accuracy from the length of the atom-pair bond.

[a)] Address all correspondence to
**trajasekharan@gmail.com** or **seshubai@gmail.com**

## I. INTRODUCTION

Since the advent of the X-ray diffraction technique, a vast amount of data has accumulated on the crystal structures adopted by intermetallic compounds. Theoretical prediction of these structures has interested many workers for a long time. Various reasons have been attributed in the literature to the choice of particular crystal structures by intermetallic compounds: they include geometrical factors, electronic factors and Brillouin zone effects, or a combination thereof (Pearson, 1972a). Many rules have been formulated using constructs such as metallic radii of atoms, valence electron concentration etc. for the purpose (Sauthhoff, 1996; Pearson, 1968; Hume Rothery & Raynor, 1954). Structural maps have been constructed using coordinates derived from empirical parameters or quantities derived from pseudo-potential calculations in an attempt to group compounds with different crystal structures in different areas on the maps (Darken & Gurry, 1953; Mooser & Pearson, 1959; Pearson, 1962; St. John & Bloch, 1974; Watson & Bennet, 1978; Zunger, 1980; Machlin & Loh, 1980; Pettifor, 1986; Villars & Hulliger, 1987). Ab initio techniques have been used to identify the correct crystal structure of an intermetallic compound from several candidate structures, and to obtain their lattice parameters ( van de Walle *et al*., 2007).

## II. THE ATOM-PAIR BOND

Miedema et al. (1975; 1980; 1992) and Boom *et al*. (1975) had proposed the following essentially empirical equation for the heat of formation $\Delta H$ of intermetallic phases:

$$\Delta H = f(c)\left[-(\Delta\phi)^2 + (Q/P)\left(\Delta N^{1/3}\right)^2 - \frac{R}{P}\right] \quad \ldots(1).$$

$\phi$ is proportional to Pauling's electronegativity; $N$ (in 'density units') was called 'electron density at the boundary of the Wigner-Seitz cells' of the elements (Miedema *et al*., 1980). $'\Delta'$

denotes the difference in two quantities for the two elements forming an alloy. According to equation (1), on a map using $|\Delta\phi|$ and $|\Delta N^{1/3}|$ as coordinates, binary metallic systems with negative and positive heats of formation would be separated by a straight line of slope $+(Q/P)^{1/2}$. The parameters $\phi$ and *N* were adjusted by Miedema et al. for 100% separation.

Rajasekharan and Girgis, ($RG$), (1983a; 1983b), had observed that binary systems with intermetallic compounds of the same crystal structure occur on a straight line on a $(\Delta\phi, \Delta N^{1/3})$ map. It has been verified that their observation is true for all the structure types in which intermetallic compounds crystallize (Rajasekharan & Seshubai, 2010a; Kameswari *et al*., 2010; Kameswari, 2008), with all the straight lines having a slope $\sim +(Q/P)^{1/2}$. The points on the map stand for binary systems; and the compositions at which the compounds occur do not have a role in deciding their coordinates. Eq. (1) predicts the signs of the heats of formation of *both liquid and solid* alloys with 100% accuracy (Boom *et al*., 1976) suggesting that long range order has no role in deciding *the signs* of the heats of formation of metallic alloys. When we consider all the compounds belonging to *all the* structure types, the only common feature among them will be the interaction on a line joining the A and B atoms, i.e. the pair-wise interaction between the A and B atoms. Hence the observations on the $RG$ map show that $\Delta\phi \propto \Delta N^{1/3}$ with the proportionality constant $\sim +(Q/P)^{1/2}$, for the A−B pair-wise interaction.

Pauling (1987a) considered charge transfer effects on the formation of intermetallic compounds. With an assumption that the A−A and B−B bonds remain unaffected in the alloying process, he arrived at the volume of an A−B ion-pair in the binary system Ca−Pb with the compounds $Ca_2Pb$, $Ca_5Pb_3$, CaPb (tP4), CaPb (cP4) and $CaPb_3$. He assumed that after equal number of Ca and Pb ions form bonds, the extra Ca and Pb atoms have the effective volumes of the elementary substances (i.e. they are not affected by the bonding process). Using the observed values of the mean atomic volume in the alloy, he then calculated the volumes of the $Ca^{+} - Pb^{-}$ ion pair in all the intermetallic compounds in the Ca−Pb system, and found them to be nearly the same. Similar observations were reported by him (Pauling, 1987a) in the case of Co−Zr and Co−Ga systems where elements of different electronic structures are involved. If we accept Pauling's' view that the A−A and B−B bonds remain unaffected in the alloying process, we might hypothesize that the heat of formation of intermetallic compounds is primarily due to covalent bonds resonating among the large number of equivalent *nearest neighbor A−B pairs*, with the longer A−B bonds contributing to a lesser extent. Hence it becomes interesting to consider the features of the bond between a single nearest neighbor A−B pair, i.e. the atom-pair bond, which can be thought of as kind of an average of the entire equivalent nearest neighbor A−B bonds.

We note that the prediction of concomitant and mutually exclusive structure types in phase diagrams is possible from the Rajasekharan-Girgis ($RG$) maps (Rajasekharan & Girgis 1983a; 1983b; Rajasekharan & Seshubai, 2010a; Kameswari *et al*., 2010; Kameswari, 2008) without any inputs regarding the compositions at which the structure types occur. The occurrence of $RG$ lines and their ability to predict concomitant structures in metallic phase diagrams is puzzling since such predictions follow from just two parameters per element: and several competing energy contributions can be expected to decide such structures. The observations were earlier unexplained. In a binary system A-B, irrespective of whether we are considering a compound occurring at AB, $AB_2$, $AB_3$ or in general any $A_mB_n$, the structures of the concomitant phases are predicted by a knowledge of which $RG$ lines or inverse $RG$ lines pass through the point corresponding to the binary system on the $(\Delta\phi,\ \Delta N^{1/3})$ map (Rajasekharan & Seshubai, 2010a). This means that the energy represented by a point on the $RG$ map (and given by the term in the bracket in Eq. (1)) is the energy of the pair-wise interaction between A and B atoms or is the A−B bond energy, which remains the same for all stoichiometry and structures involved. Also, in support of the equivalence in energy of the nearest neighbor atom-pair bonds in different compounds of the same binary system, it was shown (Rajasekharan & Seshubai, 2010a) that in *all* the 43 binary systems in which $MgCu_2$ type and $CaCu_5$ type compounds coexist, the ratio of the shortest A−B bond lengths is uniformly close to ~1.05. Equal bond lengths would mean equal bond energies. Such an effect was also demonstrated (Rajasekharan & Seshubai, 2010a) for *all* the 33 binary systems in which $MgCu_2$ and CsCl type compounds co-exist.

Metallic elements and intermetallic compounds are characterized by high ligancy (≥ 8), high symmetry and large connectivity (Laves, 1966). Pair potential calculations in simple metallic alloys and transition metal alloys have shown that the nearest neighbor pair interaction energy is more important than the further-neighbor interactions (Machlin, 1986). There are about 180 compounds which crystallize in the $MgCu_2$ type crystal structure (Pearson, 1967). In the prototype $MgCu_2$ compound (Fig. 1), each Mg atom is surrounded (Pearson, 1972d) by 12 Cu atoms at a distance of 2.922 Å. The second shortest A−B bond length in $MgCu_2$ is 4.578 Å, about 57% longer. The current view of metallic bonding is that it is an extremely delocalized communal form of covalent bonding. In a picture of resonating covalent bonds between the *A* and *B* atoms, the contribution to energy from the second nearest neighbor A-B bonds is likely to be comparatively much smaller. In the unit cells of about 50 compounds (Pearson, 1967) which crystallize in the $SiCr_3$ type crystal structure (Fig. 2), a large number of equivalent nearest neighbor A−B bonds can be observed. Another illustrative example is the $MoSi_2$ type crystal structure in which several intermetallic compounds crystallize (Pearson, 1967): a large number of equivalent nearest neighbor A−B bonds can be observed in their unit cells too (Fig. 3). The average of the equivalent nearest neighbor A-B bonds in intermetallic compounds can be thought of as the unlike atom-pair bond in a metallic alloy.

Pauling (1938; 1975; 1987; 1984) had introduced the idea that metals are covalently bonded with electrons resonating among a large number of equivalent bond positions. The large number of equivalent nearest neighbor $A - B$ bonds observed in intermetallic compounds can have its origin in the minimization of energy through resonance, and it is interesting to look for radii changes of atoms facilitating their packing in structures with high coordination numbers. Pauling had discussed the charge transfer that takes place on the unlike atom-pair bond on the formation of an alloy. He had pointed out that when a covalent bond forms, there would be an accumulation of charges on the bond near the more electronegative atom; and there has to be charge transfer from the *more* electronegative atom to the *less* electronegative atom to maintain electroneutrality (Pauling, 1950). The energy cost of such a charge transfer had not been considered in earlier approaches. In our recent work (Rajasekharan & Seshubai, 2010b), we have quantified the *positive energy* contribution to the energy of the bond from such a charge transfer. The negative term in the expression for the bond energy is due to the ionicity in the bonds. The resulting equation, with a negative and another positive term, predicts *the signs* of the heats of formation of metallic alloys accurately. Miedema's empirical constant $(Q/P)^{1/2}$ follows from the proposed model.

From the above discussion, the importance of Pauling's model for the metallic bond, and of charge transfer on the atom pair bond in the alloying of metallic elements, can be noted. Since, as per the above picture, changes during alloying are confined mostly to the nearest neighbor unlike atom-pair bond, we might expect that most the volume changes on alloying should be due to the charge transfer on the nearest neighbor unlike atom pair bond. It might be anticipated that such a charge transfer plays a crucial role in deciding the crystal structures and lattice parameters of intermetallic compounds. We verify these propositions taking some examples of well populated structure types in the subsequent sections of this paper. We have described the structure types considered in this paper only briefly; more details can be found elsewhere (Pearson, 1972). The crystallographic data are from standard sources (Pearson, 1967; Villars & Calvert, 1985; Massalski, 1990).

## II. CHARGE TRANSFER ON THE ATOM-PAIR BOND, AND SIZE CHANGES DURING ALLOYING

The binary Laves phases with an $AB_2$ composition belong to the topologically close-packed (TCP) structures (Pearson, 1972). In the group of TCP intermetallics, Laves phases constitute the largest group. They crystallize in the C14 (hP12, $MgZn_2$ prototype), the cubic C15 (cF24, $MgCu_2$ prototype), or the dihexagonal C36 (hP24, $MgNi_2$ prototype) structures. The C14, C15 and C36 structures differ only by the particular stacking of the same two-layered structural units (Hazzledine *et al*., 1993). The stability of these structure types have been proposed to be controlled by both the atomic size ratio of the $A$ and $B$ atoms, and the valence electron concentration of the Laves phases (Sauthhoff, 1996). We consider the C15 $MgCu_2$ type compounds as the first illustrative example, in this paper. Approximately 180 compounds are known which crystallize with the $MgCu_2$ type structure. An atom of diameter $D = 2R$ in the metallic element will have a different diameter $d = 2r$ in a Laves phase. The atomic size difference between the metallic (elemental) and intermetallic phases $(D - d)$ yields an apparent contraction or expansion of an atom within a Laves phase. For C15, assuming a hard sphere model, $d_{AA} = \sqrt{3}\,a/4$, $d_{BB} = \sqrt{2}\,a/4$, $d_{AB} = \sqrt{11}\,a/8$. $d_{AA}$ is the distance between two $A$ atoms in the structure, which is considered equal to the $A-A$ bond length. Similarly, $d_{BB}$ is the $B-B$ distance and $d_{AB}$ is the $A-B$ distance. $a$ is the lattice parameter. The compounds occur in the metallic radius ratio range $R_A/R_B$ = 1.05 to 1.68. The atoms forming the compounds need to adjust in size to accommodate the ideal size filling ratio $r_A/r_B$ = 1.225 in the ordered Laves phase lattice; as a result, the occurrence of Laves phases is related to the ability of the A and B atoms to change size so that the ideal ratio is approached. Mechanical stresses have been assumed to explain such size changes (Pearson, 1972; 1968). Another possible mechanism is charge transfer between the atoms on alloying as proposed by Pauling. A systematic dependence of the size changes of atoms on their electronic properties, i. e. electronegativity and valence, might be considered as supportive of the latter mechanism.

According to Pauling, charge has to be transferred from the more electronegative atom to the less electronegative atom to establish electroneutrality (Pauling, 1950). This charge transfer on the atom-pair bond has to increase with the electronegativity difference ($\Delta\chi$) between the atoms. Miedema's parameter $\phi$ is proportional to Pauling's electronegativity $\chi$ of the elements, and $\Delta\phi \propto \Delta\chi$ (Rajasekharan & Seshubai, 2010b).Thus we can expect the charge transfer on the bond to be proportional to $\Delta\phi$. (As discussed earlier, the observations on the $RG$ map show that $\Delta N^{1/3} \propto \Delta\phi$ with the proportionality constant $\sim +(Q/P)^{1/2}$ for all intermetallic compounds. We have argued elsewhere (Rajasekharan & Seshubai, 2010b) that $\Delta N^{1/3}/2$, which is proportional to $\Delta\phi$, is the charge transfer on the atom-pair bond.)

The $AsNa_3$ type compounds are hexagonal, hP8 (Pearson, 1967). Pearson had noticed that the dimensional behavior of the phases is controlled by the six Na (II) neighbors surrounding each As atom (Pearson, 1972c). Unlike the NiAs type compounds which occur with widely varying axial ratios, the 20 known phases with the $AsNa_3$ type structure are remarkable for the constancy of the axial ratio which has a value of 1.79±0.03. In Fig. 4, the $RG$ lines of $AsNa_3$ and $MgCu_2$ type structures can be seen in the first and third quadrants respectively on the ($\Delta\phi$, $\Delta N^{1/3}$) map. 19 binary systems with $AsNa_3$ type ($AB_3$) and 181 binary systems with $MgCu_2$ type ($AB_2$) compounds are considered. We note that all the $AsNa_3$ type compounds have their $\Delta\phi$ (= $(\phi_A - \phi_B)$) and $\Delta N^{1/3}$ (= $(N_A^{1/3} - N_B^{1/3})$) values positive. $MgCu_2$ type compounds have their $\Delta\phi$ (= $(\phi_A - \phi_B)$) and $\Delta N^{1/3}$ (= $(N_A^{1/3} - N_B^{1/3})$) values negative. Here, $A$ is the minority element. In Fig. 5, we plot $R_A/R_B$ versus $\Delta N^{1/3}$ for the $MgCu_2$ and $AsNa_3$ type compounds. We see that compounds exist in both systems with a wide range of $R_A/R_B$ values. No preference is shown for $R_A/R_B$ = 1.225 by the $MgCu_2$ type compounds and for any particular $R_A/R_B$ value by the $AsNa_3$ type compounds. The $R_A/R_B$ values *deviate* maximum *from* ~1.0 at $|\Delta N^{1/3}| \approx 1$, and vary to a value ~ 1.0 at $|\Delta N^{1/3}| \approx 0$. The variation of $R_A/R_B$ with $\Delta N^{1/3}$ is nearly linear. For $MgCu_2$ type compounds, $N_A^{1/3} < N_B^{1/3}$ and $\phi_A < \phi_B$. According to Pauling's model, charge is transferred from the more electronegative element to the less electronegative element, and the latter would decrease in diameter. The effect of charge transfer is to decrease the radius ratios of the atoms from the metallic radii ratio $R_A/R_B > 1$ to the ideal size filling ratio $r_A/r_B$ which is close to one. We observe from Fig. 5 that $|\Delta N^{1/3}|$, and thus the amount of charge transfer, increases proportional to the deviation of $R_A/R_B$ from 1.0. In Fig. 6, we show the change in volume on alloying as a function of $\Delta N^{1/3}$ for $MgCu_2$ type compounds; we observe that the volume change goes to zero when $\Delta N^{1/3}$ goes to zero.

A similar argument will show that the effect of charge transfer in $AsNa_3$ type compounds is to *increase* the internal radius ratio towards 1.0 by charge transfer. The greater the deviation of $R_A/R_B$ from 1, the greater is the charge transfer $\Delta N^{1/3}/2$.

Another interesting example is the case of the $MoSi_2$ type compounds. $MoSi_2$ type compounds (Pearson, 1972) are tetragonal, tI8. It is known that there are two groups of compounds of the $MoSi_2$ type structure. One group has larger axial ratio ( $c/a$) values, and the other, smaller. It can be seen from Fig. 7 that the 37 binary systems with $MoSi_2$ type compounds form two $RG$ lines with the compounds $AlTc_2$, $AlCr_2$ and $MoU_2$ as exceptions. We see from Fig. 8 in which $c/a$ are plotted versus $R_A/R_B$, that the compounds in Fig. 7 are resolved as per their axial ratios. The same compounds $AlTc_2$, $AlCr_2$ and $MoU_2$ are exceptions in both the figures. The $R_A/R_B$ values go to 1.0 on both the lines as $\Delta N^{1/3}$ goes to zero. The upper line in Fig. 7 has the same compounds as the lower line in Fig. 8: it has compounds with $R_A/R_B > 1$ and with lower $c/a$ values. The direction of charge transfer in the compounds on both the lines in Fig. 7, as given by the sign of $\Delta\phi$, is such as to alter the $R_A/R_B$ values towards the internal radius ratio $r_A/r_B \approx 1$ of the structure type. The magnitude of $\Delta\phi$ is proportionally larger when $R_A/R_B$ deviates more from 1.0. We can define $R_A - r_A$ as the change in radius of minority A atoms on alloying, where $r_A$ is the radius of the atom in the alloy. In Fig. 9, we plot $R_A - r_A$ versus $R_A/R_B$ for $MoSi_2$ type compounds. $r_A$ is calculated from the lattice parameters of the compounds as $r_A = 0.5\sqrt{\frac{a^2}{2}+\frac{c^2}{36}}$. We observe from Fig. 9, that $|R_A - r_A|$ which is nearly zero at $R_A/R_B$ =1 increases proportionately as $R_A/R_B$ deviates from one. We also observe that the increase or decrease in the sizes of the atoms is consistent with Pauling's hypothesis on which of the atoms would give up or gain electrons based on the electronegativity difference between the atoms. In Fig. 10, we plot the change in volume when $MoSi_2$ type compounds form from their constituent elements, as function of $\Delta N^{1/3}$. The change in volume goes to zero when $\Delta N^{1/3}$ goes to zero.

We consider another example: viz. the $L1_2$ ($AuCu_3$) structure type. There are about 240 compounds with this structure (Pearson, 1967), which includes $Ni_3Al$ and $TiPt_3$. It is cubic (cP4) with the minority atoms (A) at the corners of the cubic unit cell, and the majority atoms (B) ordered in the face centers. We can define $R_A - r_A$ as the change in radius of the minority atoms on alloying, where $r_A$ is the radius of atom A in the alloy. For $AuCu_3$ type structure, $r_A = a / (2\sqrt{2})$; $a$ is the lattice parameter of the compound. In Fig.11, we show a plot of $R_A - r_A$ versus $R_A/R_B$ for the 240 $AuCu_3$ type compounds. The graph can be fitted to an equation $R_A - r_A = 2.313 + 3.255\left(\frac{R_A}{R_B}\right) - 0.929\left(\frac{R_A}{R_B}\right)^2$. There are a few compounds which deviate from the curve at its ends; the deviations (~ 9 out of 240) are probably related to the large $|R_A/R_B|$ values of the deviations which make them behave like interstitial compounds. The compounds for which $\phi_A > \phi_B$ are marked by full squares (red online) and those for which $\phi_A < \phi_B$ are marked by full circles (black online). In $AuCu_3$ type compounds, the best space filling of 74% is expected (Pearson, 1972b) when $R_A/R_B = 1$. When $\phi_A > \phi_B$, we expect charge to be transferred from *A* to *B* making atom *A* larger, and atom *B* smaller. $R_A - r_A$ would be negative in such cases. We note that in such cases, the radius ratio $r_A / r_B$ after charge transfer increases to 1.0 (from $R_A/R_B$ values less than 1), thus allowing for maximum space filling. In cases where $\phi_A < \phi_B$, charge transfer is from *B* to *A* and would make the internal radius ratio to approach 1 from $R_A/R_B$ values which are larger than 1. The correlation in Fig. 12 allows the lattice parameters of new $L1_2$ compounds to be predicted with reasonable accuracy. This aspect is illustrated in Fig. 12 where we compare the calculated and observed lattice parameters of 240 compounds of the $AuCu_3$ type structure.

We therefore come to an important conclusion that the effect of charge transfer in intermetallic compounds is to bring the radius ratio of the atoms after charge transfer to a value close to 1.0. According to Pauling (1929), in ionic compounds, a coordination polyhedron of anions is formed about each cation: the cation-anion distance is determined by the radius sum $(r_c + r_a)$ and the coordination number by the radius ratio $(r_c/r_a)$. $r_c$ is the cation radius and $r_a$, the anion radius. For high coordination numbers, i.e. for CN $\geq 8$, $0.732 \leq r_c/r_a \leq 1.0$. For lower $r_c/r_a$ values, the coordination numbers of the polyhedron (CN) will be less than 8. There are a few exceptions to the above rule, but the rule is mostly obeyed. In intermetallic compounds, a ligancy >8 is generally observed (Laves, 1966). The observations of this paper show that the size change of atoms on intermetallic compound formation is correlated to the magnitude and sign of the electronegativity difference between the atoms. This size change assures that the high ligancy and coordination numbers observed in intermetallic compounds can be attained. The question of whether a particular crystal structure would be adopted by an intermetallic compound depends on whether its $\Delta\phi$ and $\Delta N^{1/3}$ are of the correct magnitude and direction on the atom-pair bond to modify the elemental $R_A/R_B$ by charge transfer to a value close to the internal $r_A/r_B$ characteristic of the structure type. The internal radius ratios, $r_A/r_B$, for crystal structures adopted by intermetallic compounds are generally close to one.

## III. THE LATTICE PARAMETERS

We will now investigate with a few more examples, the effect of change in size of the atoms on the atom-pair bond in deciding the lattice parameters of intermetallic compounds. We show a plot of $R_A - r_A$ versus $R_A/R_B$ for 181 $MgCu_2$ type ($AB_2$) compounds in Fig. 13. The $r_A$ values were obtained from the unit cell dimensions as $\frac{\sqrt{3}a}{8}$. The plot can be fitted to an equation $R_A - r_A = -1.587 + 1.534\left(\frac{R_A}{R_B}\right) - 0.168\left(\frac{R_A}{R_B}\right)^2$. There are a few compounds deviating at the ends of the curve as in the case of the $AuCu_3$ type compounds. The correlation enables one to predict accurately the $r_A$ value of a new compound knowing only its elemental $R_A/R_B$. Its lattice parameter can then be estimated to a high accuracy using the relation $a = \frac{8}{\sqrt{3}} r_A$. Fig. 14 compares the observed lattice parameters of $MgCu_2$ type compounds and the ones calculated using the present correlation. We note that in spite of considering all known binary $MgCu_2$ type compounds, including those in all elemental combinations, the difference between the experimental and calculated lattice parameter values are minimal.

In systems of lower symmetry, where the axial ratio is more or less a constant, for instance in the $AsNa_3$ type compounds, we need only one distance to decide approximately both the cell parameters. We have shown $R_A - r_A$ versus $R_A/R_B$ plot for $AsNa_3$ type compounds (hP8) with 19 representatives in Fig. 15. The $r_A$ values are calculated from the geometry of the structure as $(\frac{a}{\sqrt{3}} - \frac{\sqrt{12a^2 + c^2}}{12})$. The plot has a high regression factor and allows the lattice parameters of new hypothetical compounds with $AsNa_3$ type structure to be calculated with good accuracy, assuming an average $c/a$ value of 1.79. In Table I, we compare the experimental lattice parameters of $AsNa_3$ type compounds with those calculated from the polynomial fit in Fig. 15.

There are around 50 intermetallic compounds with A15 type ($SiCr_3$ type, cP8) crystal structure. They are an interesting group of compounds since many of them are superconducting. In Fig. 16, we plot $R_A - r_A$ versus $R_A/R_B$ for $SiCr_3$ type compounds. The $r_A$ values are obtained from the geometry of the structure as $(\sqrt{5} - 1)\, a/4$. The plot has a high regression factor and enables the lattice parameter of a new compound of the same structure type to be calculated with a maximum error of ~0.05 Å. In Table II, we compare the experimental lattice parameters of $SiCr_3$ type compounds with those calculated from the polynomial fit in Fig. 16.

Lattice parameters can be predicted for compounds belonging to several other structure types as well, from the change in radii of the atoms on charge transfer on the atom-pair bond. Lattice parameters can also be obtained for hexagonal NiAs type (hP4) compounds where the $c/a$ varies over a wide range: one can determine both $a$ and $c$ using the linear dependences of $R_A - r_A$ and $R_B - r_B$ on $R_A/R_B$.

## VI SUMMARY

The importance of the nearest neighbor atom-pair bond in the alloying of metals has been discussed. According to Pauling, charge transfer to maintain electroneutrality is an important characteristic of the bond. The energy expenditure in transferring charges from the more electronegative atom to the less electronegative atom had been quantified by us earlier and an equation for the energy of the atom-pair bond was derived. That equation yielded Miedema's empirical constant $\sqrt{Q/P}$ , and could predict the signs of the heats of formation of metallic alloys accurately.

We have demonstrated in this paper that the change in size of the atoms on alloy formation is strongly correlated to the electronic properties of metallic elements. Such a size change is, therefore, unlikely to be decided by mechanical stresses as advocated earlier. The direction and magnitude of size change of atoms on alloying supports Pauling's picture of charge transfer on the metallic atom-pair bond. It is proportional to the electronegativity difference between the atoms.

In intermetallic compounds there is a shortage of electrons in comparison with the available bond positions. There is scope to lower energy through resonance if several equivalent bonds are formed in their unit cells. For the atoms in intermetallic compounds to have high ligancy, and for them to have coordination polyhedra of high symmetry of one type of partially ionised atoms around the other type, the radius ratios of the ionized atoms have to be close to 1.0. We show in this paper that the electronegativity and valence differences on the atom-pair bond in alloys are in such a direction as to bring the metallic radius ratio $R_A/R_B$ by charge transfer to a value $r_A/r_B$ ~ 1.0. If the deviation of elemental radius ratio $(\frac{R_A}{R_B})$ from the internal radius ratio in the intermetallic compound $(r_A/r_B)$ is more, the charge transfer is proportionally larger. The role of stresses and Brillouin zone effects in making minor changes to the $R_A/R_B$ values is not ruled out, but the role of such effects appear to be minor.

We have demonstrated in this paper that the change in radii of the atoms on the nearest neighbor atom-pair bond decides the unit cell dimensions of intermetallic compounds. Knowing just the elemental $R_A/R_B$ value, one can predict the lattice parameters of a new hypothetical intermetallic compound to good accuracy.

The work reported in the present paper, and in other recent papers discussing the atom-pair bond in metallic alloys, is interesting from the view point of predicting the properties of the metallic systems for use in the laboratory/industry without recourse to time consuming calculations. It is also interesting to note that the picture of pair-wise interaction in metals and alloys has the potential to be the starting point in techniques such as Monte Carlo or Molecular Dynamics simulations of metallic systems.

## VII ACKNOWLEDGEMENTS

TR thanks DMRL, Hyderabad, India for permission to publish this paper.

**Table I**. The experimental lattice parameters of $AsNa_3$ type compounds along with those calculated using the correlation in Fig. 15. The differences between the experimental and calculated values are also given. The lattice parameters are in Å.

| No. | Cpd. | *a* -exp | *a*-cal | *c*-exp | *c*-cal | Diff in *a* | Diff. in *c* |
|---|---|---|---|---|---|---|---|
| 1 | $HgMg_3$ | 4.87 | 4.88 | 8.66 | 8.69 | 0.01 | -0.03 |
| 2 | $IrAl_3$ | 4.25 | 4.13 | 7.76 | 7.36 | -0.12 | 0.40 |
| 3 | $AuMg_3$ | 4.84 | 4.64 | 8.46 | 8.27 | -0.20 | 0.19 |
| 4 | $AsLi_3$ | 4.38 | 4.42 | 7.80 | 7.87 | 0.04 | -0.07 |
| 5 | $BiNa_3$ | 5.46 | 5.65 | 9.67 | 10.07 | 0.19 | -0.40 |
| 6 | $PtMg_3$ | 4.58 | 4.51 | 8.32 | 8.04 | -0.07 | 0.28 |
| 7 | $PdMg_3$ | 4.61 | 4.48 | 8.41 | 7.98 | -0.13 | 0.43 |
| 8 | $IrMg_3$ | 4.55 | 4.46 | 8.23 | 7.95 | -0.09 | 0.28 |
| 9 | $SbNa_3$ | 5.37 | 5.42 | 9.52 | 9.66 | 0.05 | -0.14 |
| 10 | $PLi_3$ | 4.27 | 4.24 | 7.59 | 7.55 | -0.04 | 0.05 |
| 11 | $AsNa_3$ | 5.1 | 5.07 | 9.00 | 9.04 | -0.03 | -0.04 |
| 12 | $BiK_3$ | 6.19 | 6.34 | 10.96 | 11.29 | 0.15 | -0.33 |
| 13 | $BiRb_3$ | 6.42 | 6.52 | 11.46 | 11.62 | 0.10 | -0.16 |
| 14 | $SbK_3$ | 6.04 | 6.13 | 10.71 | 10.93 | 0.09 | -0.21 |
| 15 | $PNa_3$ | 4.99 | 4.92 | 8.82 | 8.77 | -0.07 | 0.05 |
| 16 | $SbRb_3$ | 6.25 | 6.32 | 11.12 | 11.26 | 0.07 | -0.14 |
| 17 | $AsK_3$ | 5.79 | 5.80 | 10.24 | 10.34 | 0.01 | -0.10 |
| 18 | $AsRb_3$ | 6.05 | 5.99 | 10.73 | 10.67 | -0.06 | 0.06 |
| 19 | $PK_3$ | 5.69 | 5.65 | 10.05 | 10.07 | -0.04 | -0.02 |

**Table II**. The lattice parameters of 50 $SiCr_3$ type compounds calculated using the polynomial fit in Fig. 16 are compared with the experimental values. All elemental combinations with $SiCr_3$ type compounds are considered. The difference between the calculated and experimental values is mostly ≤ 0.05Å. Lattice parameter values are in Å.

| No. | Cpd. | *a*-cal | *a*-exp | Diff. | No. | Cpd. | *a*-cal | *a*-exp | Diff. |
|---|---|---|---|---|---|---|---|---|---|
| 1 | $SnV_3$ | 4.98 | 4.96 | -0.02 | 26 | $SiCr_3$ | 4.56 | 4.64 | 0.08 |
| 2 | $SbV_3$ | 4.93 | 4.93 | 0.00 | 27 | $IrV_3$ | 4.79 | 4.77 | -0.02 |
| 3 | $SnMo_3$ | 5.09 | 5.09 | 0.00 | 28 | $GaMo_3$ | 4.94 | 4.93 | -0.01 |
| 4 | $InNb_3$ | 5.29 | 5.31 | 0.02 | 29 | $AUNb_3$ | 5.21 | 5.15 | -0.05 |
| 5 | $SnNb_3$ | 5.29 | 5.27 | -0.02 | 30 | $RhV_3$ | 4.78 | 4.76 | -0.02 |
| 6 | $AuV_3$ | 4.88 | 4.82 | -0.06 | 31 | $AuTi_3$ | 5.10 | 5.18 | 0.08 |
| 7 | $SbNb_3$ | 5.26 | 5.25 | -0.02 | 32 | $GeMo_3$ | 4.94 | 4.92 | -0.03 |
| 8 | $SnTa_3$ | 5.28 | 5.35 | 0.07 | 33 | $SiV_3$ | 4.73 | 4.76 | 0.03 |
| 9 | $GaCr_3$ | 4.65 | 4.68 | 0.03 | 34 | $HgZr_3$ | 5.56 | 5.58 | 0.02 |
| 10 | $SbTi_3$ | 5.22 | 5.27 | 0.05 | 35 | $AlNb_3$ | 5.19 | 5.13 | -0.05 |
| 11 | $AsCr_3$ | 4.62 | 4.67 | 0.05 | 36 | $IrMo_3$ | 4.97 | 4.91 | -0.05 |
| 12 | $PtCr_3$ | 4.71 | 4.67 | -0.04 | 37 | $OsMo_3$ | 4.97 | 4.91 | -0.06 |
| 13 | $HgTi_3$ | 5.19 | 5.25 | 0.07 | 38 | $GaNb_3$ | 5.18 | 5.12 | -0.05 |
| 14 | $AlV_3$ | 4.81 | 4.80 | -0.01 | 39 | $PtNb_3$ | 5.14 | 5.12 | -0.02 |
| 15 | $GeCr_3$ | 4.63 | 4.66 | 0.03 | 40 | $SiMo_3$ | 4.90 | 4.90 | 0.00 |
| 16 | $GaV_3$ | 4.82 | 4.79 | -0.03 | 41 | $PtTi_3$ | 5.03 | 5.14 | 0.11 |
| 17 | $IrCr_3$ | 4.68 | 4.65 | -0.03 | 42 | $CoV_3$ | 4.68 | 4.74 | 0.07 |
| 18 | $OsCr_3$ | 4.68 | 4.65 | -0.03 | 43 | $IrNb_3$ | 5.13 | 5.10 | -0.03 |
| 19 | $PtV_3$ | 4.82 | 4.78 | -0.03 | 44 | $NiV_3$ | 4.71 | 4.74 | 0.03 |
| 20 | $AsV_3$ | 4.75 | 4.78 | 0.03 | 45 | $IrTi_3$ | 5.01 | 5.13 | 0.12 |
| 21 | $RhCr_3$ | 4.67 | 4.65 | -0.03 | 46 | $OsNb_3$ | 5.14 | 5.10 | -0.04 |
| 22 | $Rucr_3$ | 4.68 | 4.65 | -0.03 | 47 | $RhNb_3$ | 5.12 | 5.10 | -0.02 |
| 23 | $AlMo_3$ | 4.95 | 4.95 | 0.00 | 48 | $AuZr_3$ | 5.49 | 5.51 | 0.02 |
| 24 | $PdV_3$ | 4.83 | 4.77 | -0.05 | 49 | $SiNb_3$ | 5.16 | 5.09 | -0.07 |
| 25 | $GeV_3$ | 4.78 | 4.77 | -0.01 | 50 | $BeMo_3$ | 4.89 | 4.88 | -0.01 |

Fig. 1

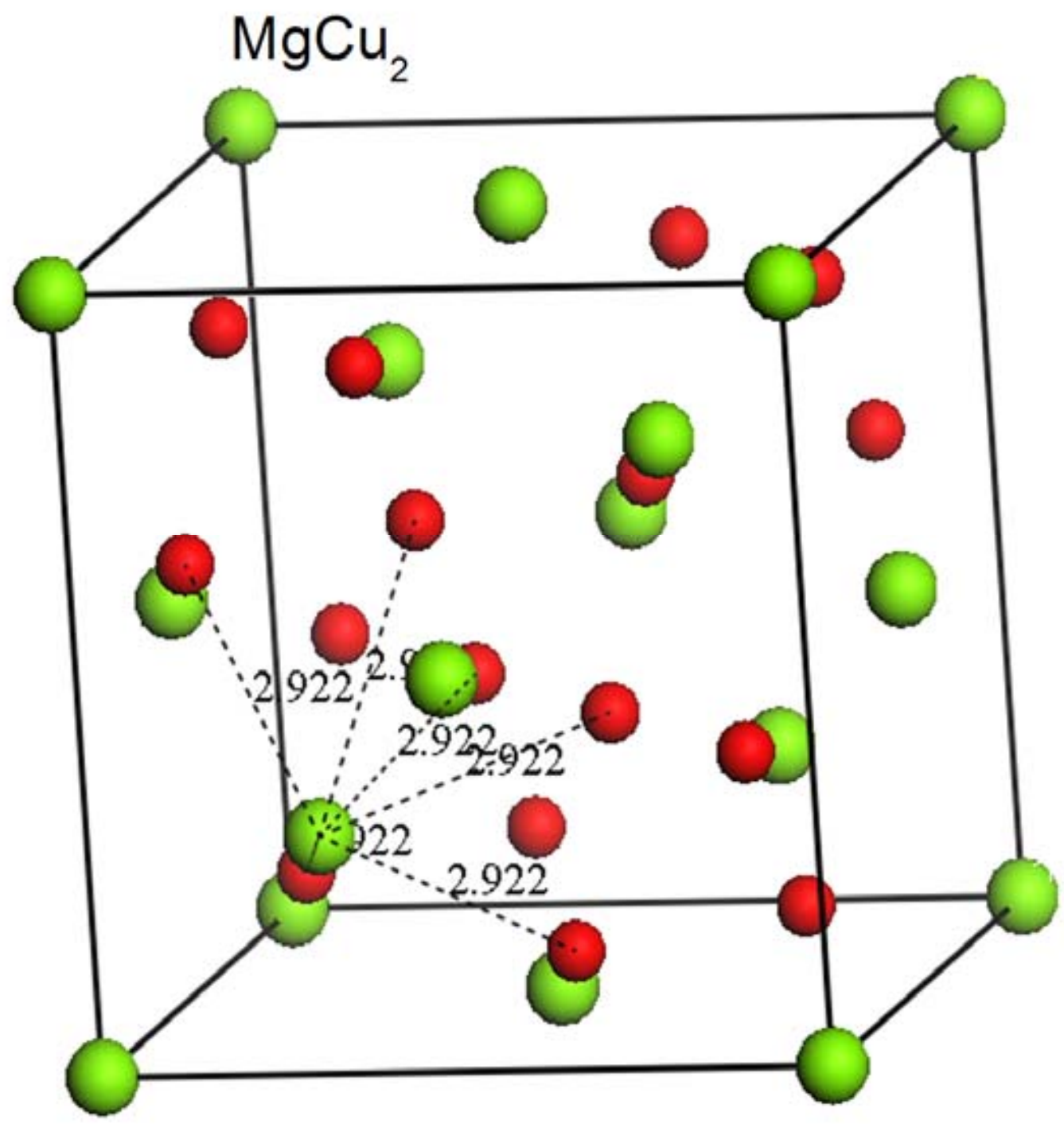

**Fig. 1** The unit cell of $MgCu_2$ ($AB_2$, cF24) compound. *A* atoms are shown larger in diameter (green online). There are 12 bonds of equal length to the *B* atoms from each *A* atom. In the compound $MgCu_2$, the Mg-Cu nearest neighbor distance is 2.922 Å. The next nearest neighbor Mg-Cu distance is 57% more. Assuming a hard sphere model, $d_{AA} = \sqrt{3}\,a/4$, $d_{BB} = \sqrt{2}\,a/4$, $d_{AB} = \sqrt{11}\,a/8$. The distance between the *A* atoms at (0, 0, 0) and (¼, ¼, ¼) in the unit cell is $2r_A = \frac{\sqrt{3}a}{4}$.

Fig. 2

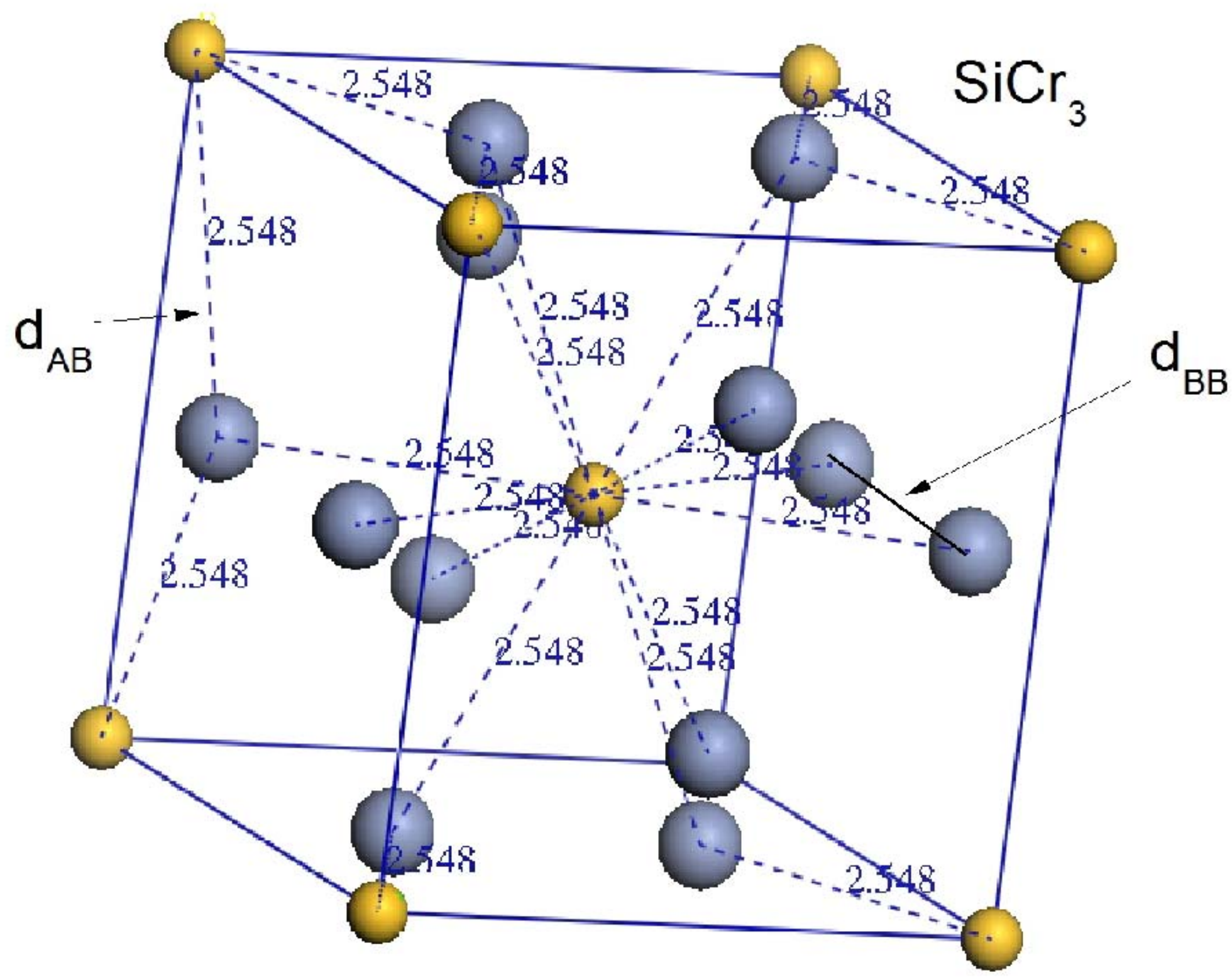

**Fig. 2** The unit cell of $SiCr_3$ type ($AB_3$, cP8) compounds. An abundance of $A-B$ bonds of equal length can be seen in the prototype $SiCr_3$ compound. $d_{AB}$ = $\sqrt{5}a/4$, $d_{BB}$ =$a/2$, $r_A$ = $(\sqrt{5}-1)a/4$.

Fig. 3

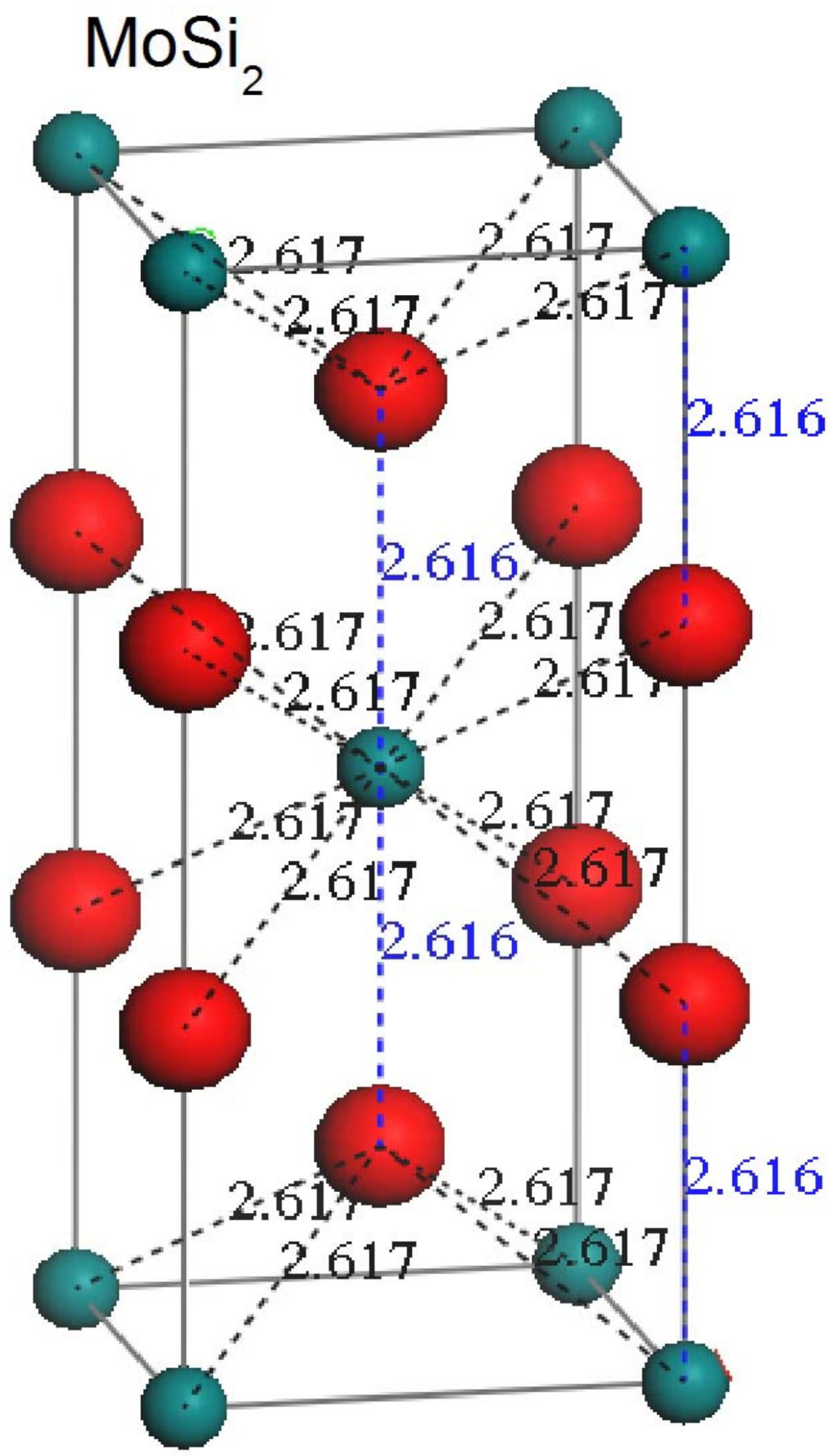

**Fig. 3** The unit cell of $MoSi_2$ type ($AB_2$, tI8) compounds. The $A$ atoms are shown smaller (green online) and the B atoms are shown bigger (red online). Several $A - B$ bonds of equal length can be seen in the prototype $MoSi_2$ compound.

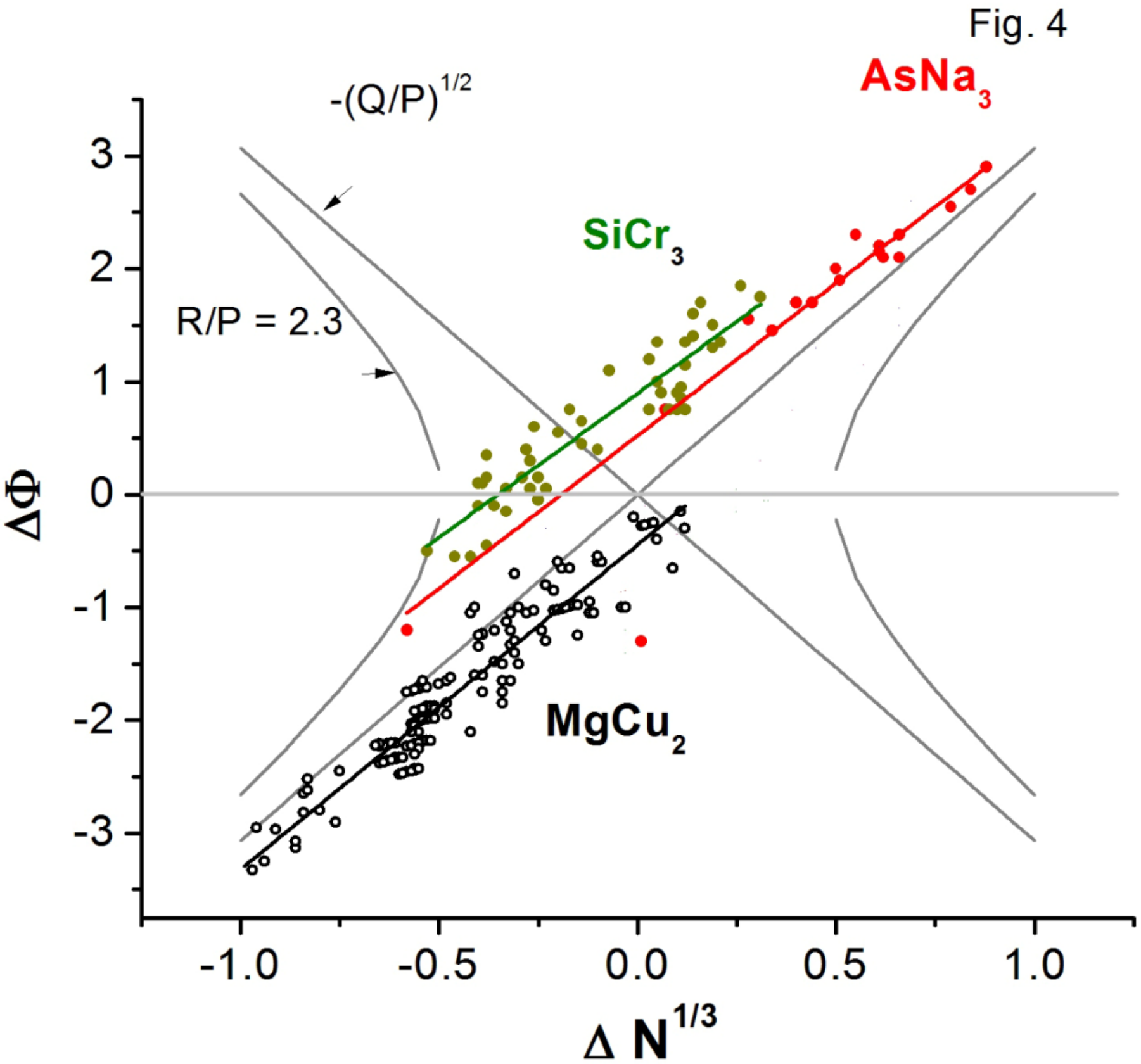

**Fig. 4** The $RG$ lines of $MgCu_2$, $AsNa_3$ and $SiCr_3$ crystal structure types. The binary systems in which compounds of those crystal structures occur are plotted on a $(\Delta\phi, \Delta N^{1/3})$ map. $\Delta\phi = (\phi_A - \phi_B)$ and $\Delta N^{1/3} = (N_A^{1/3} - N_B^{1/3})$, where $A$ is the minority element. Binary systems with compounds belonging to each structure type fall on a straight line on the map. From the overlap and lack of overlap between the lines, one can predict concomitant and mutually exclusive structure types in binary systems (Rajasekharan & Seshubai, 2010a). Most of the compounds of the $MgCu_2$ type crystal structure have negative $\Delta\phi$, and most of the compounds with $SiCr_3$ and $AsNa_3$ type structure have positive $\Delta\phi$.

Fig. 5

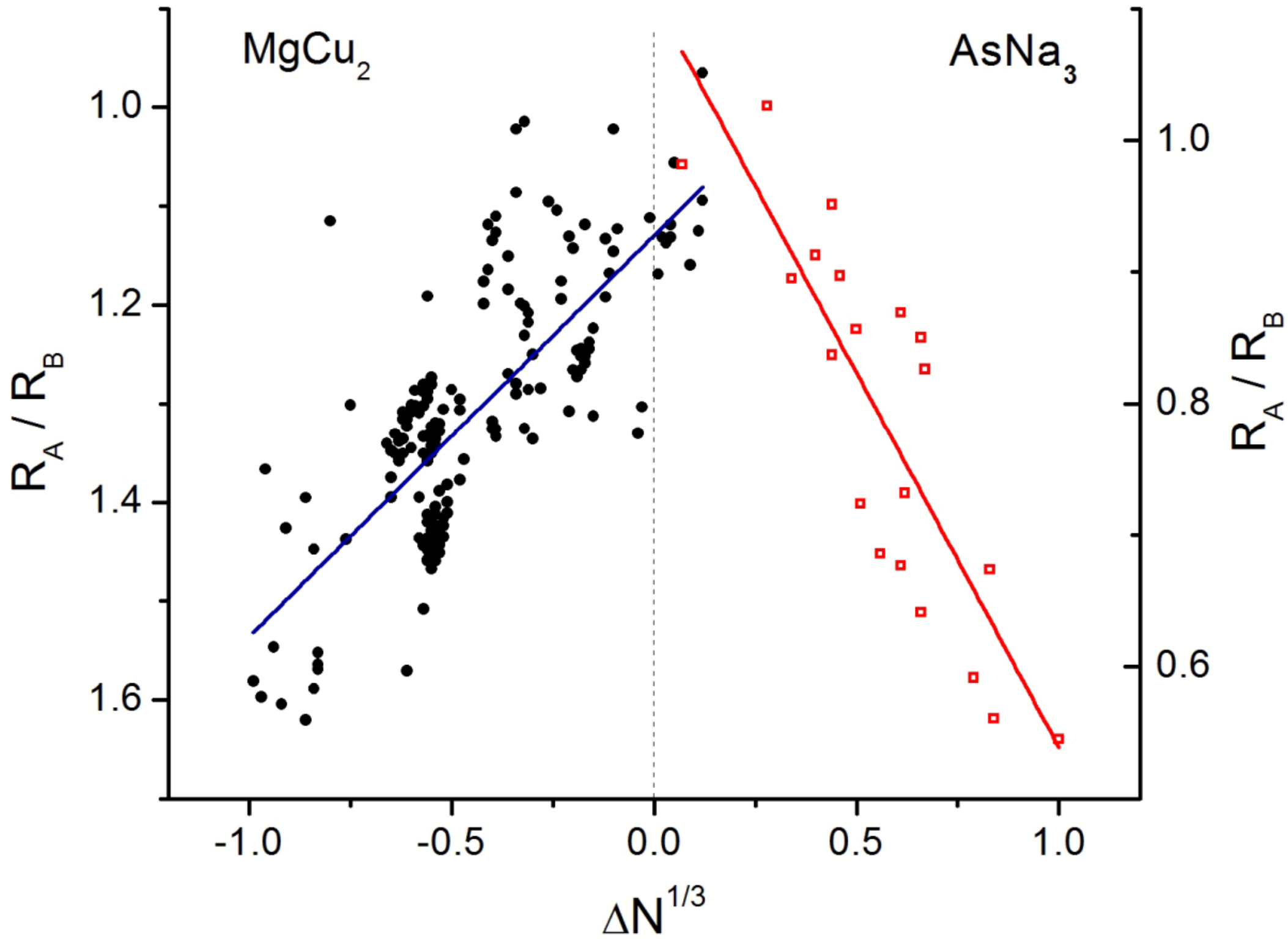

**Fig. 5** $R_A/R_B$ is plotted versus $\Delta N^{1/3}$ for the $MgCu_2$ and $AsNa_3$ type compounds. Compounds exist in both systems in a wide range of $R_A/R_B$ values. No preference is shown for any particular metallic radii ratios. $A$ is the minority element. $R_A$ and $R_B$ are the radii of the atoms in the elemental state. $\Delta N^{1/3} = N_A^{1/3} - N_B^{1/3}$ is negative for $MgCu_2$ type compounds and so is $\Delta\phi = \phi_A - \phi_B$. $\Delta\phi \propto \Delta N^{1/3}$ for the compounds as can be seen from Fig. 4. Direction of charge transfer is from the more electronegative atom to the less electronegative atom (Pauling, 1950), i. e. from $B$ to $A$ and the $A$ atoms become smaller due the enhanced attraction by the positive charges at the core. The radius ratios of the atoms shift from $R_A/R_B$ which are in a range above 1.0 to values suitable for ideal packing in the $MgCu_2$ type structure. Maximum change in radii is for compounds with $R_A/R_B$ >> 1 and with $\Delta N^{1/3} | \approx 1$; the minimum change is when $R_A/R_B \sim 1.0$ at $|\Delta N^{1/3}| \approx 0$. The variation of $R_A/R_B$ with $\Delta N^{1/3}$ is nearly linear. A similar argument will show that radii ratios will shift from $R_A/R_B$ values which are in a range less than 1.0 for $AsNa_3$ type compounds, to values close to 1.0 by charge transfer.

Fig. 6

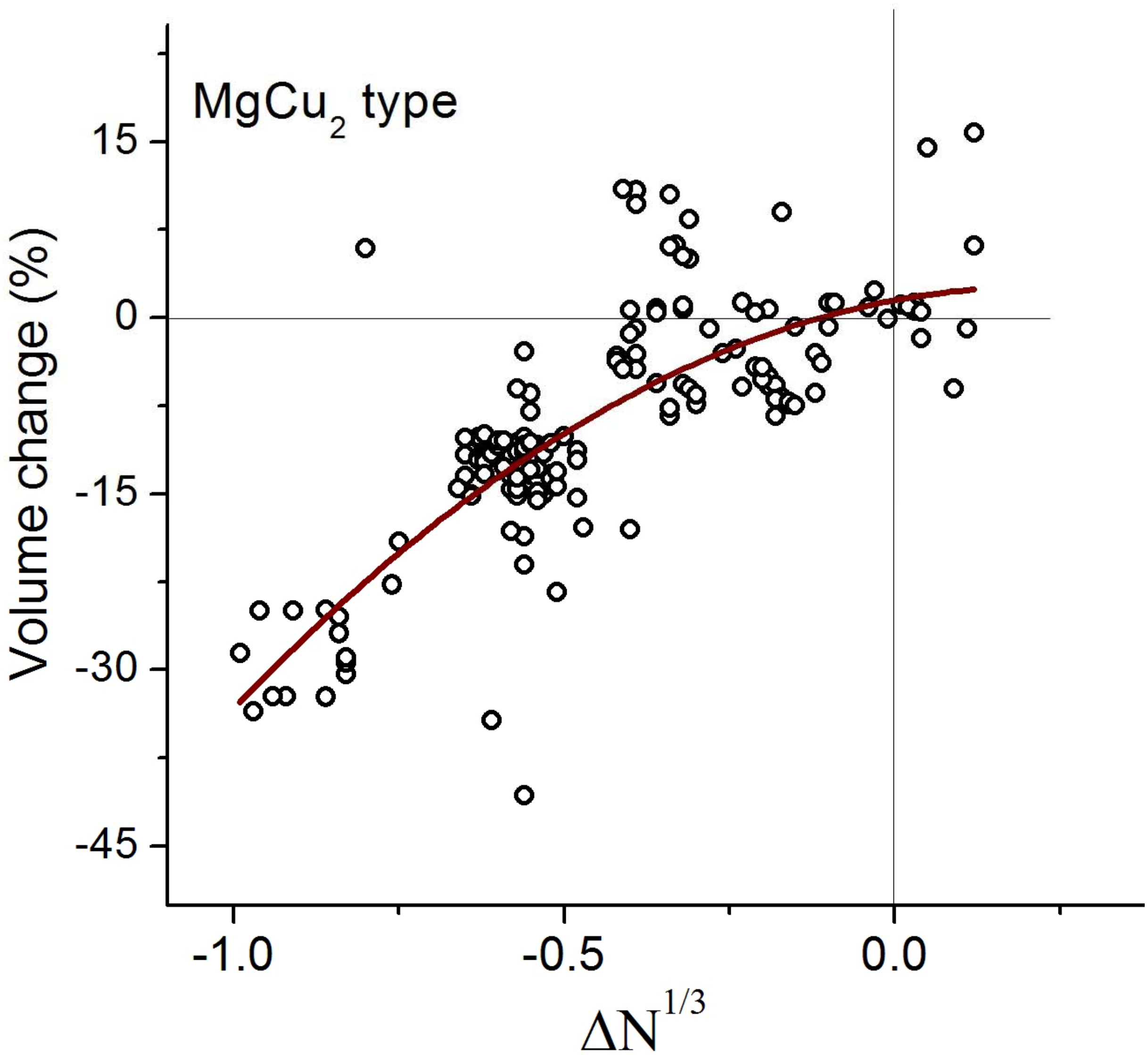

**Fig. 6** Percentage change in volume on alloying as a function of $\Delta N^{1/3}$ for $MgCu_2$ type compounds. The volume per molecule estimated from the molar volumes of the elements, is subtracted from the volume per molecule computed from the unit cell volumes of the compounds, and expressed as a percentage relative to the volume per molecule before alloying. It can be observed that the change in volume goes to zero when $\Delta N^{1/3}$ goes to zero. This observation supports the idea that $\Delta N^{1/3}$ is proportional to the charge transfer on alloying.

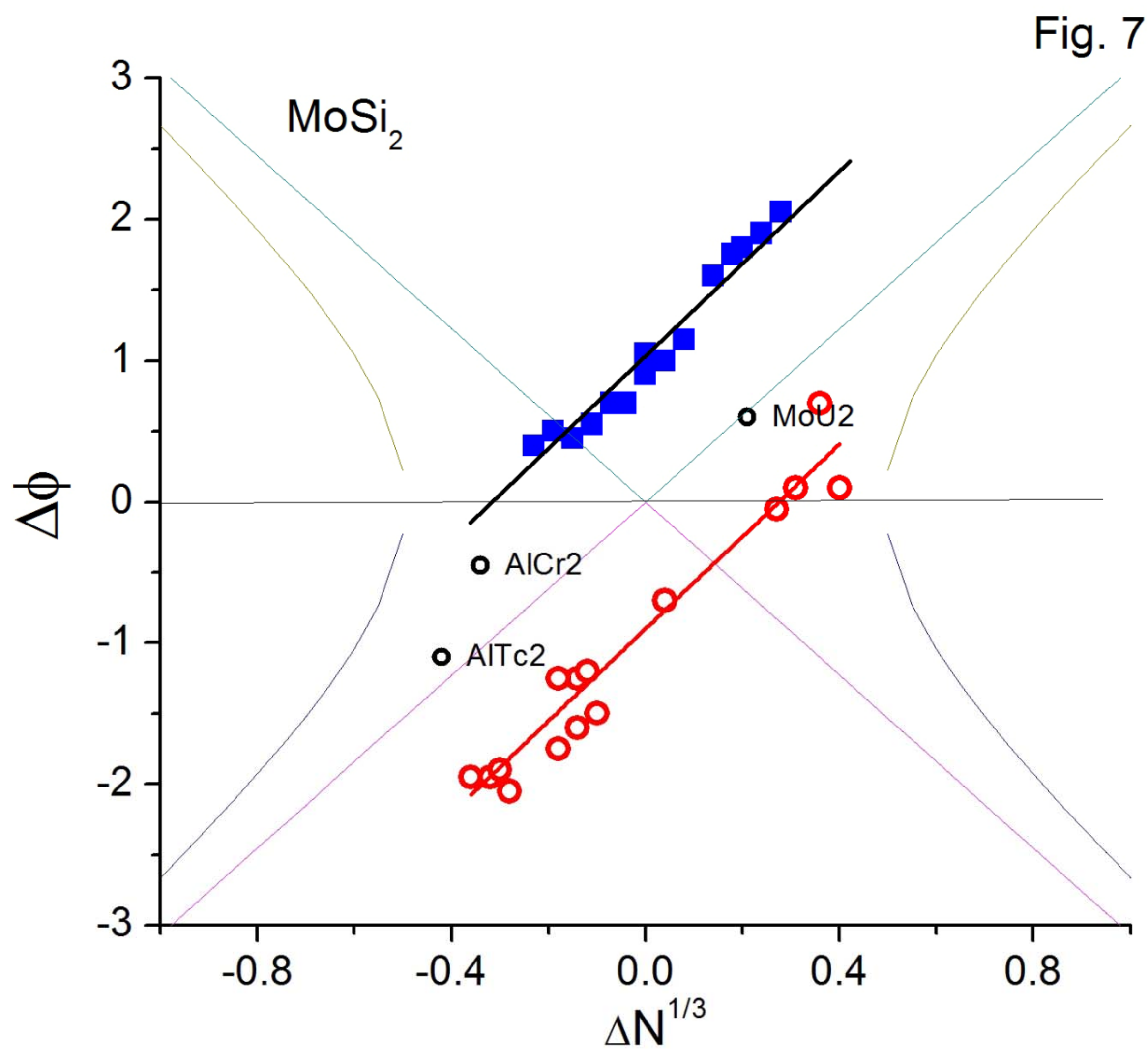

**Fig. 7** Binary systems with $MoSi_2$ type compounds occur on two straight lines on the ($\Delta\phi$, $\Delta N^{1/3}$) map. $AlCr_2$, $AlTc_2$ and $MoU_2$ (open circles) are exceptions. $A$ is the minority element. Full squares (blue online) represents compounds with $\Delta\phi = \phi_A - \phi_B$ positive and open circles (red online) represent compounds with $\Delta\phi = \phi_A - \phi_B$ negative.

Fig. 8

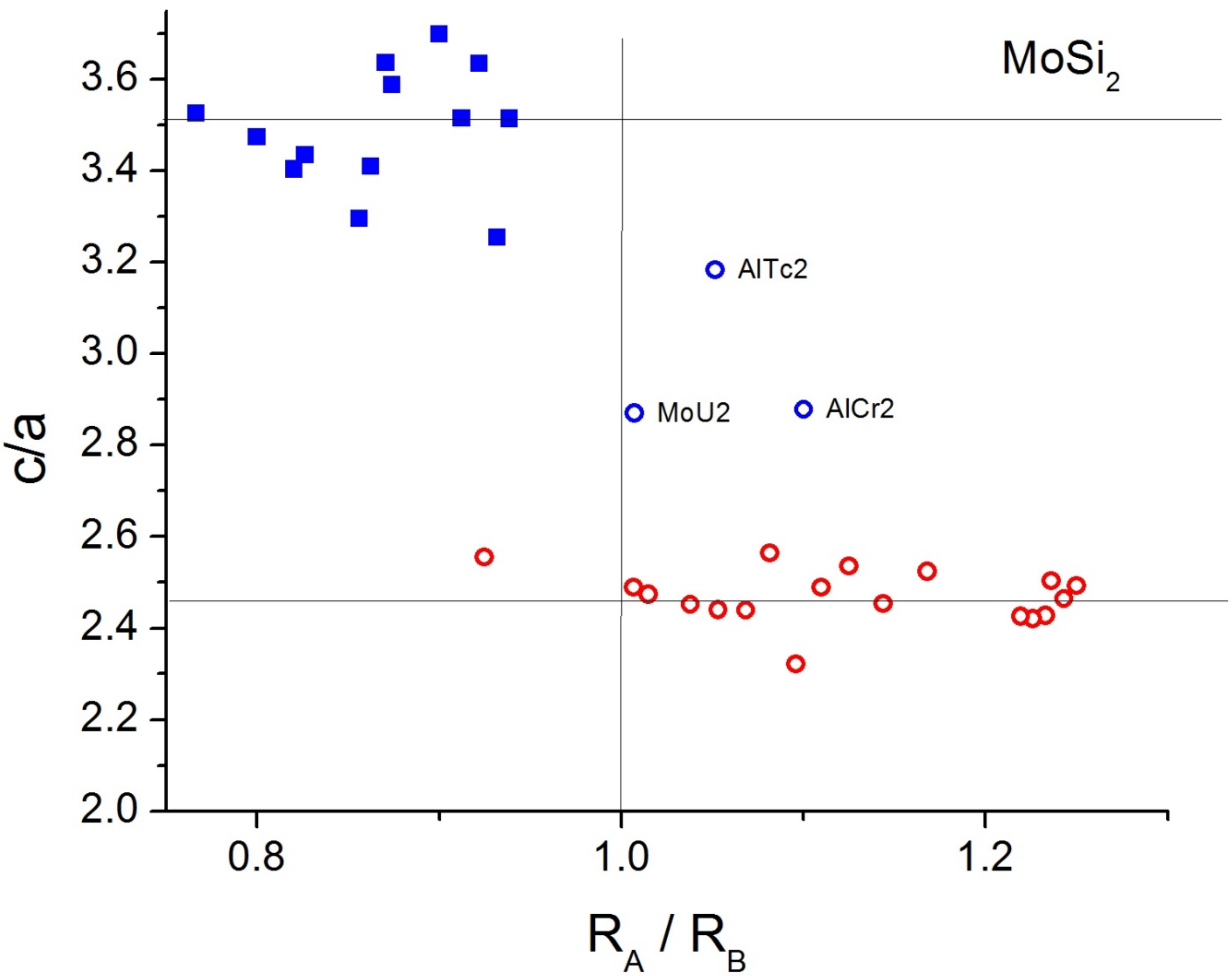

**Fig. 8** On a plot of $c/a$ versus $R_A/R_B$, the $MoSi_2$ type compounds form two groups. $AlCr_2$, $AlTc_2$ and $MoU_2$ (open circles) which were exceptions in Fig. 7 are exceptions in this figure too. The compounds with $\Delta\phi = \phi_A - \phi_B$ positive (full squares, blue online) have higher $c/a$ values and have $R_A/R_B$ values less than 1.0. The direction of charge transfer (Pauling, 1950) is appropriate to shift the radius ratio of the atoms to 1.0 in the structure. The compounds with $\Delta\phi = \phi_A - \phi_B$ negative (open circles, red online) have lower $c/a$ values and have $R_A/R_B$ values more than 1.0. The direction of charge transfer is appropriate to shift the radius ratio of the atoms to 1.0 in the structure.

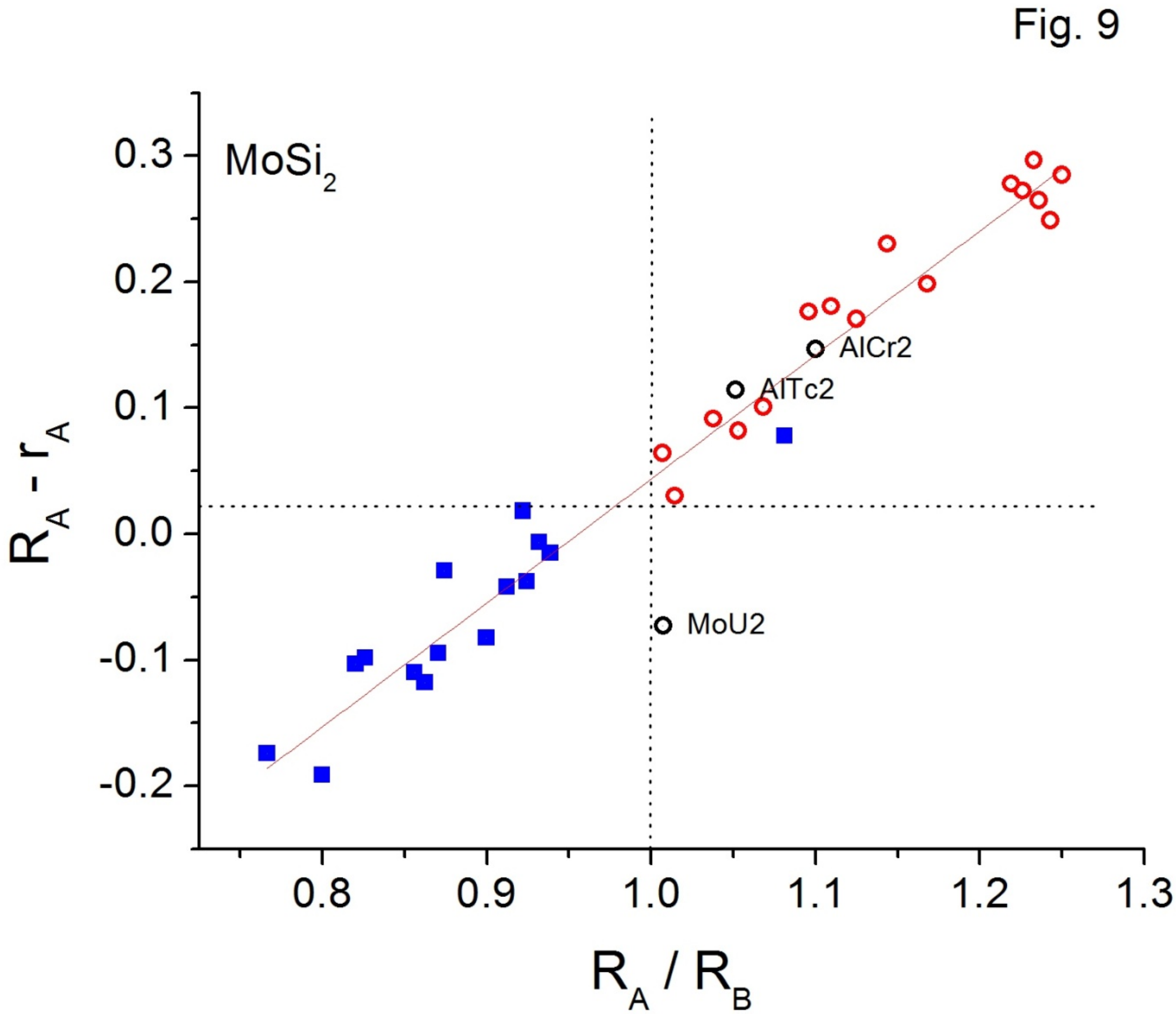

**Fig. 9** Change in radius of the minority atom on alloying i.e. $R_A - r_A$ is plotted versus $R_A/R_B$ for $MoSi_2$ type compounds. $r_A$ is the radius of the $A$ atom in the structure and is calculated from the lattice parameters as $0.5\sqrt{\frac{a^2}{2}+\frac{c^2}{36}}$ . The plot is linear and $R_A - r_A$ is nearly zero when $R_A/R_B = 1$. The compounds with $\Delta\phi = \phi_A - \phi_B$ positive (full squares, blue online) have $R_A - r_A$ negative, and $|R_A - r_A|$ increases with the deviation of $R_A/R_B$ from 1. The compounds with $\phi_A - \phi_B$ negative (open circles, red online) have $r_A$ decreasing with increasing $R_A/R_B$.

Fig. 10

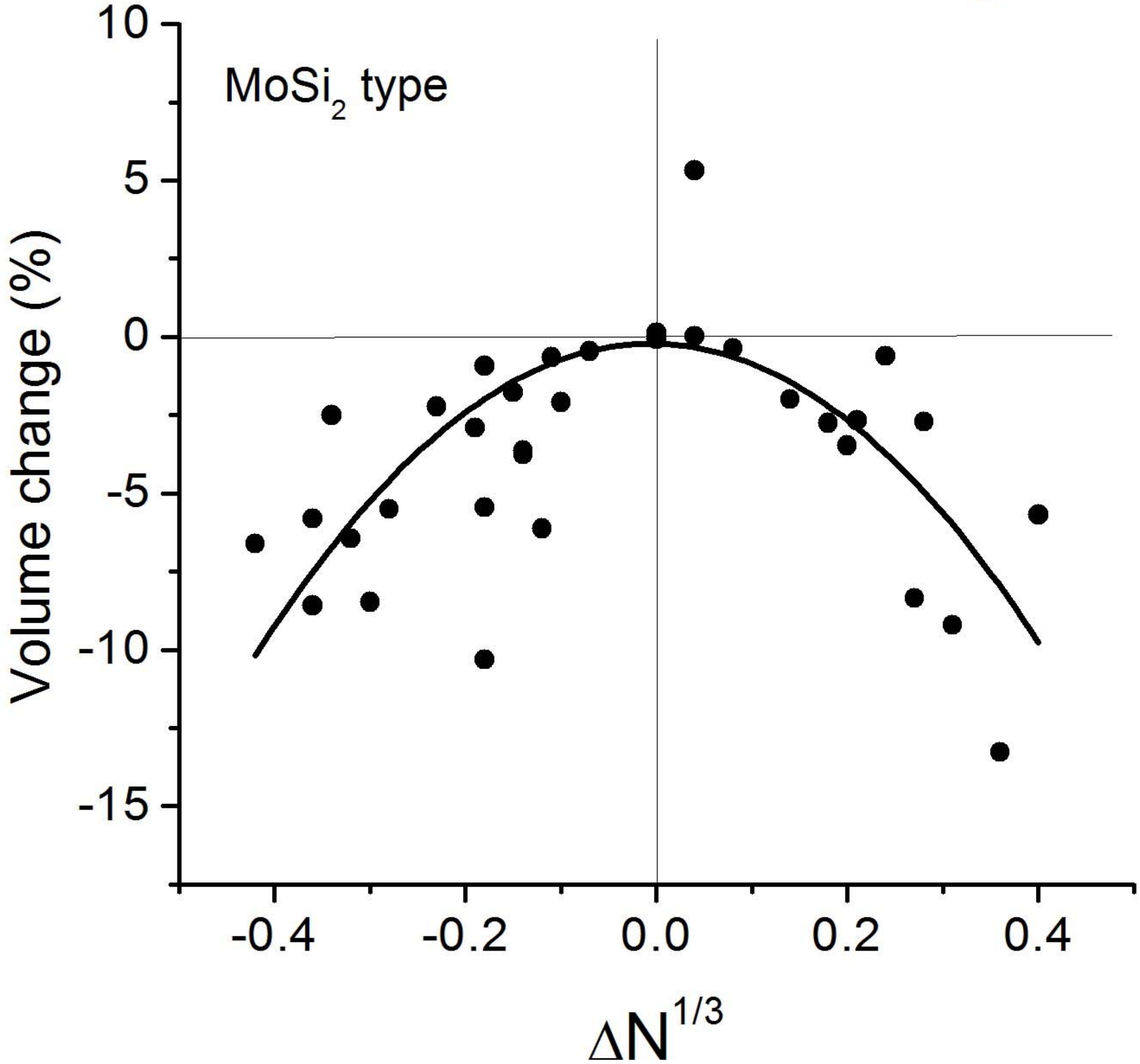

**Fig. 10** Percentage change in volume on alloying as a function of $\Delta N^{1/3}$ for $MoSi_2$ type compounds. The volume per molecule estimated from the molar volumes of the elements is subtracted from the volume per molecule computed from the unit cell volumes of the compounds, and expressed as a percentage relative to the volume per molecule before alloying. It can be observed that the change in volume goes to zero when $\Delta N^{1/3}$ goes to zero. This observation supports the idea that $\Delta N^{1/3}$ is proportional to the charge transfer on alloying.

Fig. 11

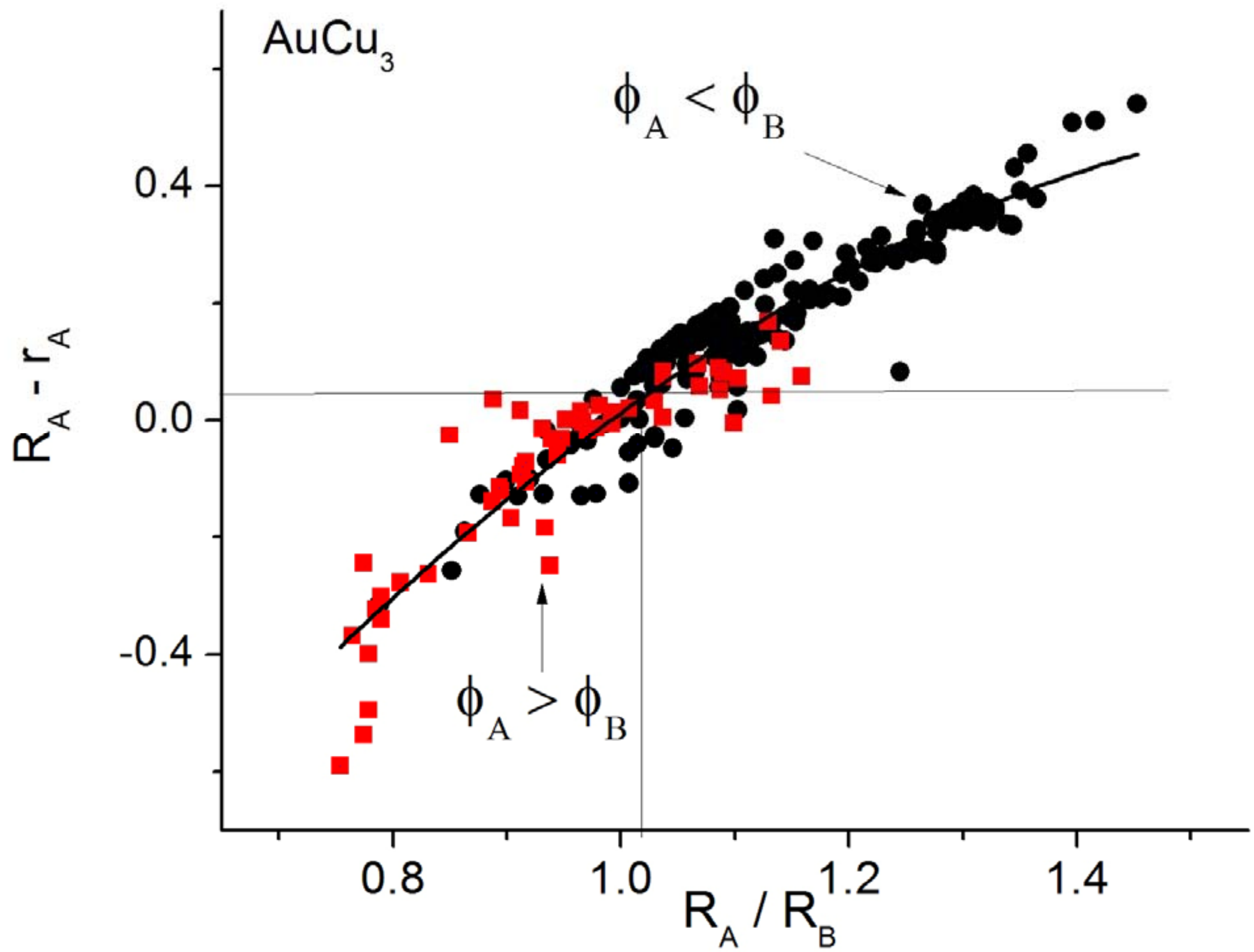

**Fig. 11** A plot of the change in radius of the minority atom on alloying $R_A - r_A$ versus $R_A/R_B$ for 240 $AuCu_3$ type compounds. $r_A$ is the radius of the $A$ atom in the structure and is calculated from the lattice parameter as $a\,/\,(2\sqrt{2})$. The points can be fitted to an equation $R_A - r_A = 2.313 + 3.255\left(\frac{R_A}{R_B}\right) - 0.929\left(\frac{R_A}{R_B}\right)^2$. $R_A - r_A$ is nearly zero when $R_A/R_B = 1$. The compounds with $\phi_A > \phi_B$ (full squares, red online) have $R_A - r_A$ negative, and those with $\phi_A < \phi_B$ (full circles, black online) have $R_A - r_A$ positive. The change in size of the atoms is in accordance with the hypothesis that the more electronegative atom transfers electrons to the less electronegative one on alloying.

Fig. 12

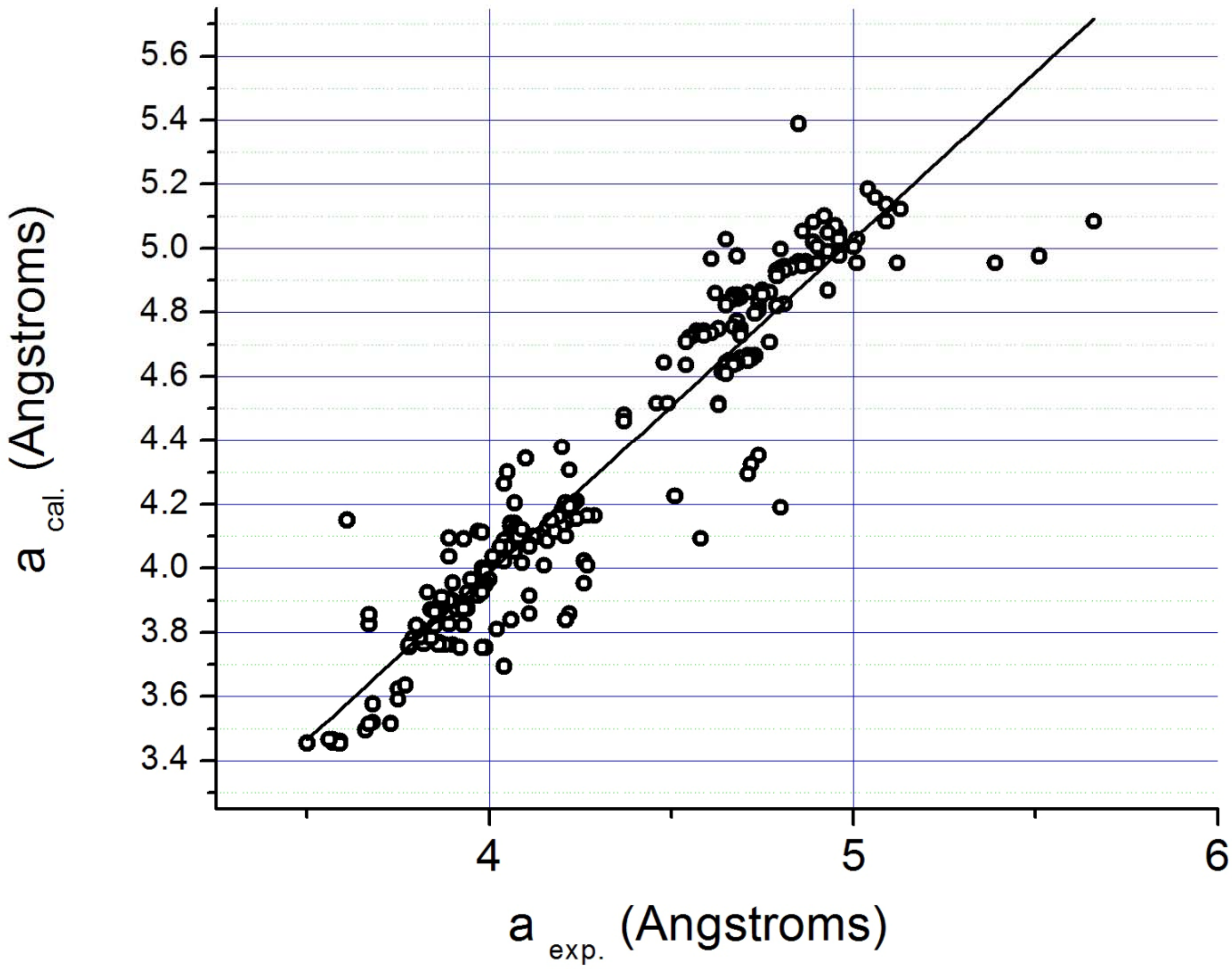

**Fig. 12** A plot of the lattice parameters of 240 $AuCu_3$ type compounds calculated using the correlation observed in Fig. 11, versus the experimentally observed values. The lattice parameters of the compounds depend on the change in radius of the atoms on charge transfer. One observes that lattice parameters of new compounds can be predicted with reasonable accuracy starting with the elemental radius ratios.

Fig. 13

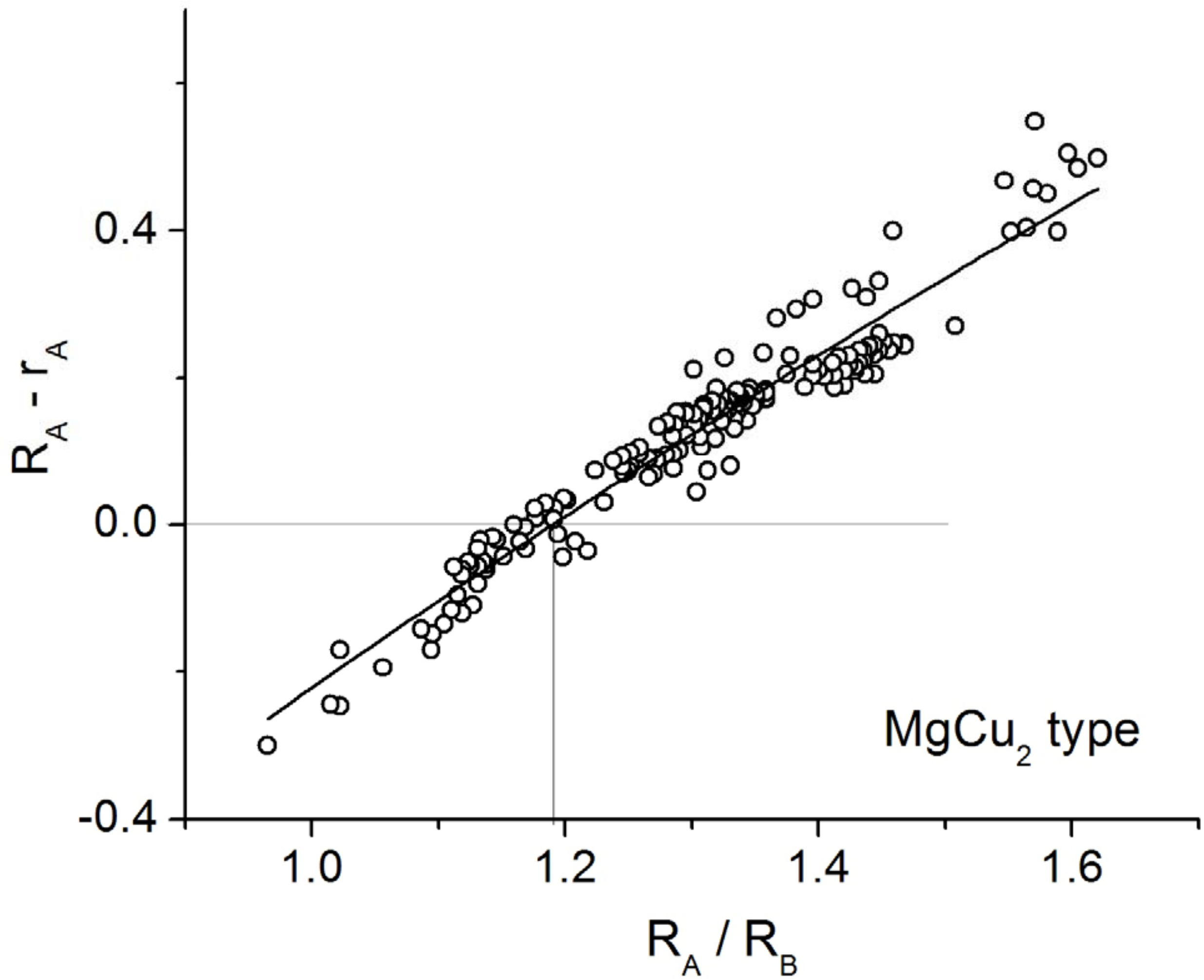

**Fig. 13** A plot of the change in radius of the minority atom on alloying $R_A - r_A$ versus $R_A/R_B$ for 181 $MgCu_2$ type compounds. $r_A$ is the radius of the *A* atom in the structure and is calculated from the lattice parameter as $\frac{\sqrt{3}a}{8}$. The points can be fitted to an equation $R_A - r_A = -1.587 + 1.534\left(\frac{R_A}{R_B}\right) - 0.168\left(\frac{R_A}{R_B}\right)^2$. A systematic deviation is observed for the points at the end of the line as in the case of the $AuCu_3$ type compounds in Fig. 11. $R_A - r_A$ is nearly zero when $\frac{R_A}{R_B} \sim 1.2$, close to the internal radius ratio of 1.225 proposed by Laves (1966).

Fig. 14

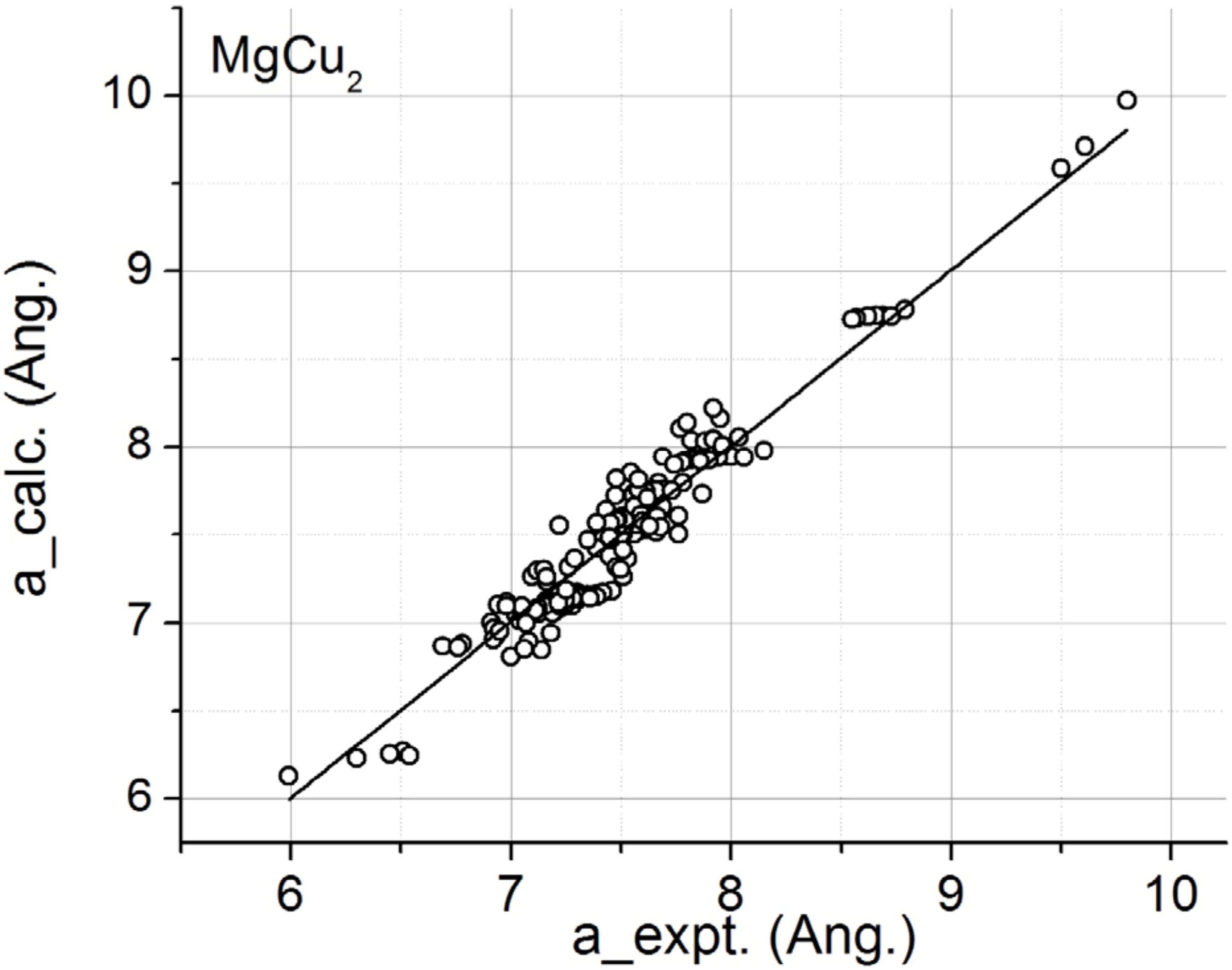

**Fig. 14** A plot of the lattice parameters 181 $MgCu_2$ type compounds calculated using the correlation observed in Fig. 11, versus the experimentally observed values. The lattice parameters of the compounds depend on the change in radius of the atoms on charge transfer. One observes that lattice parameters of new compounds can be predicted with reasonable accuracy starting with the elemental radius ratios.

Fig. 15

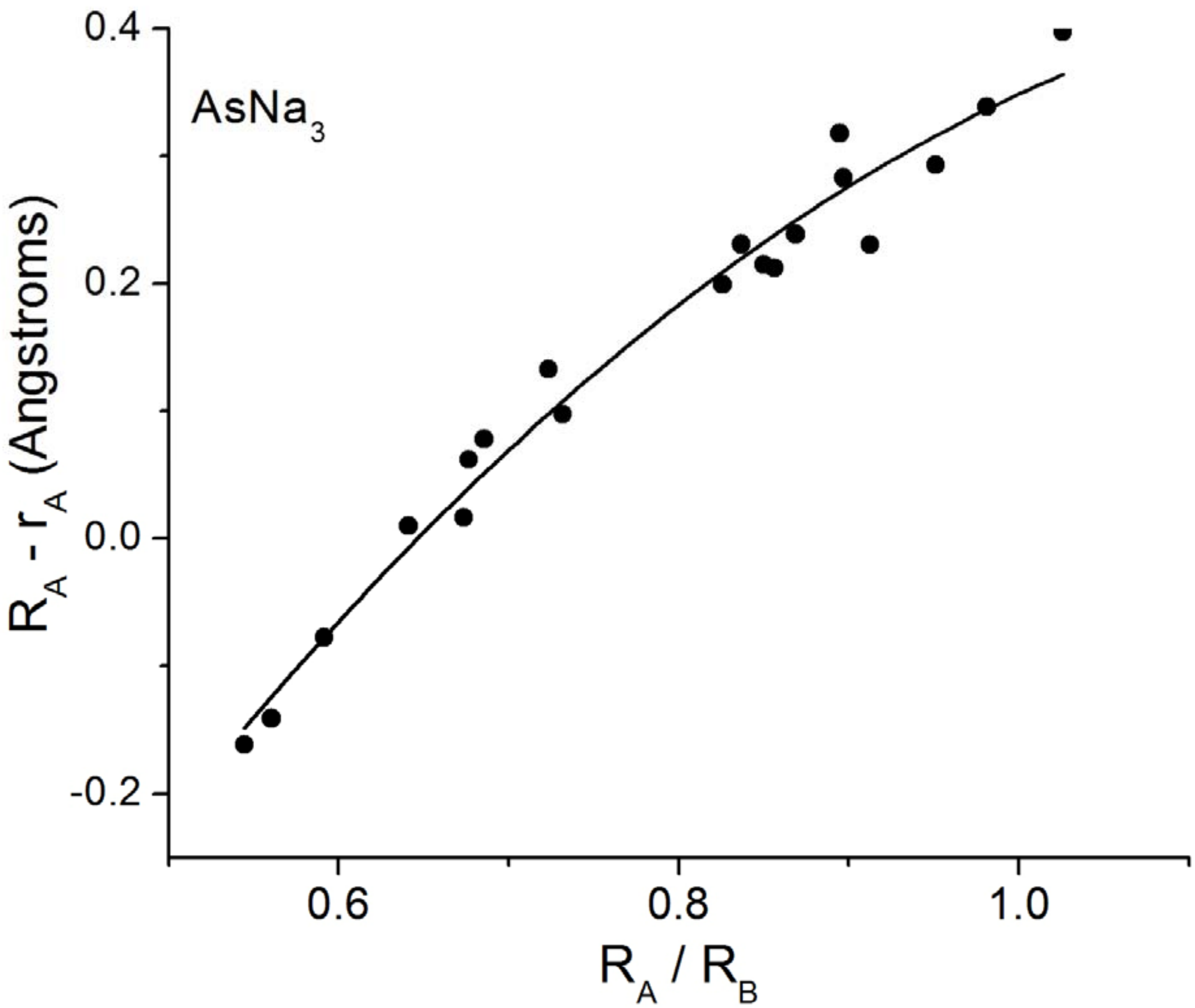

**Fig. 15** A plot of the change in radius $R_A - r_A$ of the minority atoms on alloying, versus $R_A/R_B$ for 19 $AsNa_3$ type compounds. $r_A$ is the radius of the $A$ atom in the structure and is calculated from the lattice parameters as $r_A = (\frac{a}{\sqrt{3}} - \frac{\sqrt{12a^2 + c^2}}{12})$. The points can be fitted to an equation $R_A - r_A = -1.308 + 2.693\left(\frac{R_A}{R_B}\right) - 1.037\left(\frac{R_A}{R_B}\right)^2$. Lattice parameters of new compounds of the same structure type can estimated with good accuracy from the observed correlation, as demonstrated in Table I.

Fig. 16

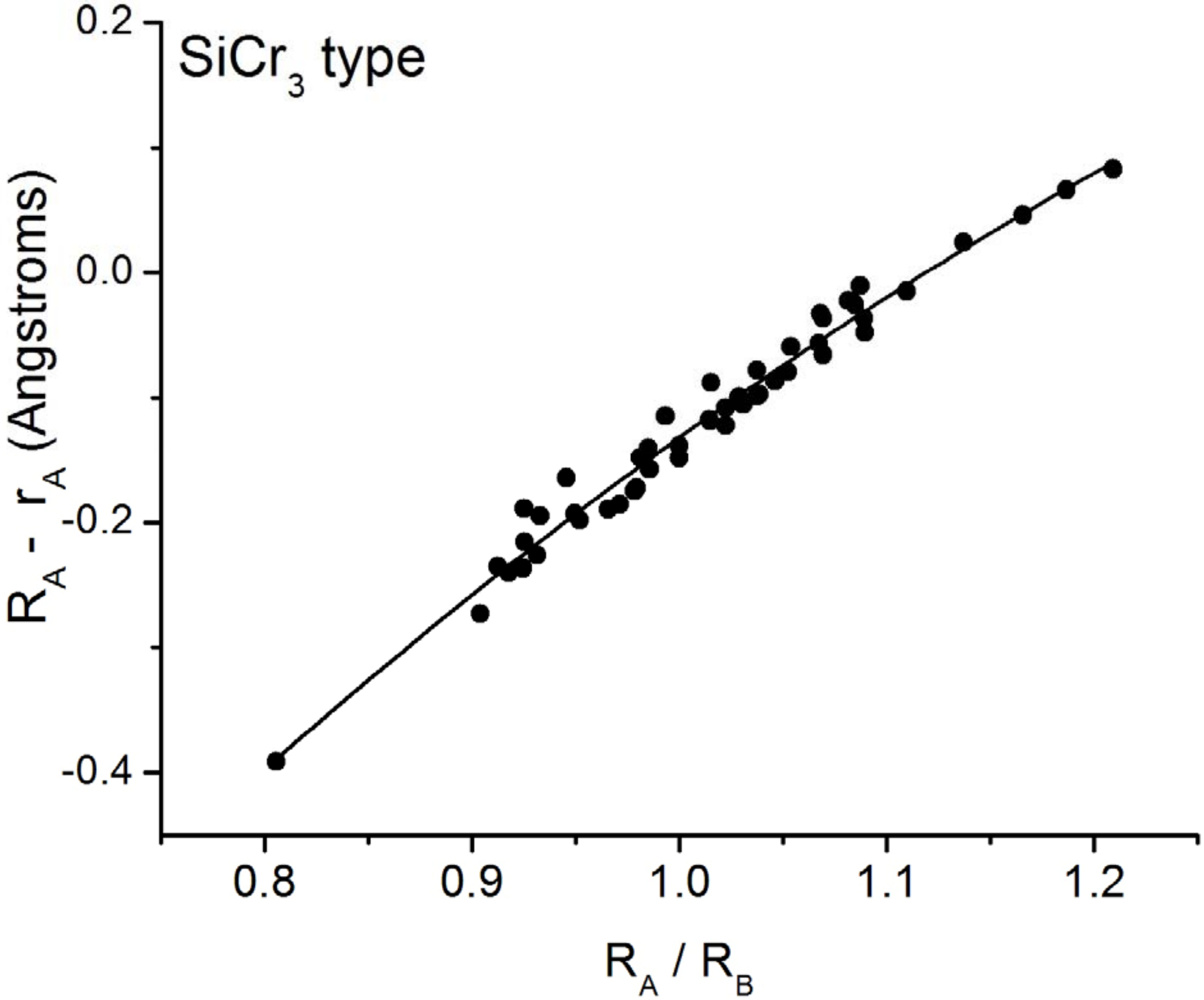

**Fig. 16** A plot of the change in radius $R_A - r_A$ of the minority atom on alloying, versus $R_A/R_B$ for 50 $SiCr_3$ type compounds. $r_A$ is the radius of the $A$ atom in the structure and is calculated from the lattice parameter as $(\sqrt{5}-1)\,a/4$. The points can be fitted to an equation $R_A - r_A = -2.002 + 2.544\left(\frac{R_A}{R_B}\right) - 0.675\left(\frac{R_A}{R_B}\right)^2$. Lattice parameters of new compounds of the same structure type can estimated with good accuracy from the observed correlation, as demonstrated in Table II.

*FIGURE CAPTIONS*

**Fig. 1** The unit cell of $MgCu_2$ ($AB_2$, cF24) compound. $A$ atoms are shown larger in diameter (green online). There are 12 bonds of equal length to the $B$ atoms from each $A$ atom. In the compound $MgCu_2$, the Mg-Cu nearest neighbor distance is 2.922 Å. The next nearest neighbor Mg-Cu distance is 57% more. Assuming a hard sphere model, $d_{AA} = \sqrt{3}\,a/4$, $d_{BB} = \sqrt{2}\,a/4$, $d_{AB} = \sqrt{11}\,a/8$. The distance between the $A$ atoms at (0, 0, 0) and (¼, ¼, ¼) in the unit cell is $2r_A = \frac{\sqrt{3}a}{4}$.

**Fig. 2** The unit cell of $SiCr_3$ type ($AB_3$, cP8) compounds. An abundance of $A - B$ bonds of equal length can be seen in the prototype $SiCr_3$ compound. $d_{AB} = \sqrt{5}a/4$, $d_{BB} = a/2$, $r_A = (\sqrt{5} - 1)a/4$.

**Fig. 3** The unit cell of $MoSi_2$ type ($AB_2$, tI8) compounds. The $A$ atoms are shown smaller (green online) and the B atoms are shown bigger (red online). Several $A - B$ bonds of equal length can be seen in the prototype $MoSi_2$ compound.

**Fig. 4** The $RG$ lines of $MgCu_2$, $AsNa_3$ and $SiCr_3$ crystal structure types. The binary systems in which compounds of those crystal structures occur are plotted on a ($\Delta\phi$, $\Delta N^{1/3}$) map. $\Delta\phi = (\phi_A - \phi_B)$ and $\Delta N^{1/3} = (N_A^{1/3} - N_B^{1/3})$, where $A$ is the minority element. Binary systems with compounds belonging to each structure type fall on a straight line on the map. From the overlap and lack of overlap between the lines, one can predict concomitant and mutually exclusive structure types in binary systems (Rajasekharan & Seshubai, 2010a). Most of the compounds of the $MgCu_2$ type crystal structure have negative $\Delta\phi$, and most of the compounds with $SiCr_3$ and $AsNa_3$ type structure have positive $\Delta\phi$.

**Fig. 5** $R_A/R_B$ is plotted versus $\Delta N^{1/3}$ for the $MgCu_2$ and $AsNa_3$ type compounds. Compounds exist in both systems in a wide range of $R_A/R_B$ values. No preference is shown for any particular metallic radii ratios. *A* is the minority element. $R_A$ and $R_B$ are the radii of the atoms in the elemental state. $\Delta N^{1/3} = N_A^{1/3} - N_B^{1/3}$ is negative for $MgCu_2$ type compounds and so is $\Delta\phi = \phi_A - \phi_B$. $\Delta\phi \propto \Delta N^{1/3}$ for the compounds as can be seen from Fig. 4. Direction of charge transfer is from the more electronegative atom to the less electronegative atom (Pauling, 1950), i. e. from *B* to *A* and the *A* atoms become smaller due the enhanced attraction by the positive charges at the core. The radius ratios of the atoms shift from $R_A/R_B$ which are in a range above 1.0 to values suitable for ideal packing in the $MgCu_2$ type structure. Maximum change in radii is for compounds with $R_A/R_B >> 1$ and with $\Delta N^{1/3}| \approx 1$; the minimum change is when $R_A/R_B \sim 1.0$ at $|\Delta N^{1/3}| \approx 0$. The variation of $R_A/R_B$ with $\Delta N^{1/3}$ is nearly linear. A similar argument will show that radii ratios will shift from $R_A/R_B$ values which are in a range less than 1.0 for $AsNa_3$ type compounds, to values close to 1.0 by charge transfer.

**Fig. 6** Percentage change in volume on alloying as a function of $\Delta N^{1/3}$ for $MgCu_2$ type compounds. The volume per molecule estimated from the molar volumes of the elements, is subtracted from the volume per molecule computed from the unit cell volumes of the compounds, and expressed as a percentage relative to the volume per molecule before alloying. It can be observed that the change in volume goes to zero when $\Delta N^{1/3}$ goes to zero. This observation supports the idea that $\Delta N^{1/3}$ is proportional to the charge transfer on alloying.

**Fig. 7** Binary systems with $MoSi_2$ type compounds occur on two straight lines on the ($\Delta\phi$, $\Delta N^{1/3}$) map. $AlCr_2$, $AlTc_2$ and $MoU_2$ (open circles) are exceptions. *A* is the minority element.

Full squares (blue online) represents compounds with $\Delta\phi = \phi_A - \phi_B$ positive and open circles (red online) represent compounds with $\Delta\phi = \phi_A - \phi_B$ negative.

**Fig. 8** On a plot of $c/a$ versus $R_A/R_B$, the $MoSi_2$ type compounds form two groups. $AlCr_2$, $AlTc_2$ and $MoU_2$ (open circles) which were exceptions in Fig. 7 are exceptions in this figure too. The compounds with $\Delta\phi = \phi_A - \phi_B$ positive (full squares, blue online) have higher $c/a$ values and have $R_A/R_B$ values less than 1.0. The direction of charge transfer (Pauling, 1950) is appropriate to shift the radius ratio of the atoms to 1.0 in the structure. The compounds with $\Delta\phi = \phi_A - \phi_B$ negative (open circles, red online) have lower $c/a$ values and have $R_A/R_B$ values more than 1.0. The direction of charge transfer is appropriate to shift the radius ratio of the atoms to 1.0 in the structure.

**Fig. 9** Change in radius of the minority atom on alloying i.e. $R_A - r_A$ is plotted versus $R_A/R_B$ for $MoSi_2$ type compounds. $r_A$ is the radius of the *A* atom in the structure and is calculated from the lattice parameters as $0.5\sqrt{\frac{a^2}{2}+\frac{c^2}{36}}$ . The plot is linear and $R_A - r_A$ is nearly zero when $R_A/R_B = 1$. The compounds with $\Delta\phi = \phi_A - \phi_B$ positive (full squares, blue online) have $R_A - r_A$ negative, and $|R_A - r_A|$ increases with the deviation of $R_A/R_B$ from 1. The compounds with $\phi_A - \phi_B$ negative (open circles, red online) have $r_A$ decreasing with increasing $R_A/R_B$.

**Fig. 10** Percentage change in volume on alloying as a function of $\Delta N^{1/3}$ for $MoSi_2$ type compounds. The volume per molecule estimated from the molar volumes of the elements is subtracted from the volume per molecule computed from the unit cell volumes of the compounds, and expressed as a percentage relative to the volume per molecule before

alloying. It can be observed that the change in volume goes to zero when $\Delta N^{1/3}$ goes to zero. This observation supports the idea that $\Delta N^{1/3}$ is proportional to the charge transfer on alloying.

**Fig. 11** A plot of the change in radius of the minority atom on alloying $R_A - r_A$ versus $R_A/R_B$ for 240 $AuCu_3$ type compounds. $r_A$ is the radius of the *A* atom in the structure and is calculated from the lattice parameter as $a\,/\,(2\sqrt{2})$. The points can be fitted to an equation $R_A - r_A = 2.313 + 3.255\left(\frac{R_A}{R_B}\right) - 0.929\left(\frac{R_A}{R_B}\right)^2$. $R_A - r_A$ is nearly zero when $R_A/R_B = 1$. The compounds with $\phi_A > \phi_B$ (full squares, red online) have $R_A - r_A$ negative, and those with $\phi_A < \phi_B$ (full circles, black online) have $R_A - r_A$ positive. The change in size of the atoms is in accordance with the hypothesis that the more electronegative atom transfers electrons to the less electronegative one on alloying.

**Fig. 12** A plot of the lattice parameters of 240 $AuCu_3$ type compounds calculated using the correlation observed in Fig. 11, versus the experimentally observed values. The lattice parameters of the compounds depend on the change in radius of the atoms on charge transfer. One observes that lattice parameters of new compounds can be predicted with reasonable accuracy starting with the elemental radius ratios.

**Fig. 13** A plot of the change in radius of the minority atom on alloying $R_A - r_A$ versus $R_A/R_B$ for 181 $MgCu_2$ type compounds. $r_A$ is the radius of the *A* atom in the structure and is calculated from the lattice parameter as $\frac{\sqrt{3}a}{8}$. The points can be fitted to an equation $R_A - r_A = -1.587 + 1.534\left(\frac{R_A}{R_B}\right) - 0.168\left(\frac{R_A}{R_B}\right)^2$. A systematic deviation is observed for the points at the end

of the line as in the case of the $AuCu_3$ type compounds in Fig. 11. $R_A - r_A$ is nearly zero when $\frac{R_A}{R_B} \sim 1.2$, close to the internal radius ratio of 1.225 proposed by Laves (1966).

**Fig. 14** A plot of the lattice parameters 181 $MgCu_2$ type compounds calculated using the correlation observed in Fig. 11, versus the experimentally observed values. The lattice parameters of the compounds depend on the change in radius of the atoms on charge transfer. One observes that lattice parameters of new compounds can be predicted with reasonable accuracy starting with the elemental radius ratios.

**Fig. 15** A plot of the change in radius $R_A - r_A$ of the minority atoms on alloying, versus $R_A/R_B$ for 19 $AsNa_3$ type compounds. $r_A$ is the radius of the $A$ atom in the structure and is calculated from the lattice parameters as $r_A = (\frac{a}{\sqrt{3}} - \frac{\sqrt{12a^2 + c^2}}{12})$. The points can be fitted to an equation $R_A - r_A = -1.308 + 2.693 \left(\frac{R_A}{R_B}\right) - 1.037 \left(\frac{R_A}{R_B}\right)^2$. Lattice parameters of new compounds of the same structure type can estimated with good accuracy from the observed correlation, as demonstrated in Table I.

**Fig. 16** A plot of the change in radius $R_A - r_A$ of the minority atom on alloying, versus $R_A/R_B$ for 50 $SiCr_3$ type compounds. $r_A$ is the radius of the $A$ atom in the structure and is calculated from the lattice parameter as $(\sqrt{5} - 1)\,a/4$. The points can be fitted to an equation $R_A - r_A = -2.002 + 2.544 \left(\frac{R_A}{R_B}\right) - 0.675 \left(\frac{R_A}{R_B}\right)^2$. Lattice parameters of new compounds of the same structure type can estimated with good accuracy from the observed correlation, as demonstrated in Table II.